\documentclass[traditabstract]{aa} % for the abstract without structuration 
\pdfoutput=1
\usepackage{graphicx}
\usepackage{color}
\usepackage{url}
\usepackage{amssymb}
\usepackage{amsmath}
\usepackage{txfonts}
\usepackage{multirow}
\usepackage{lscape}
\usepackage{float}
\usepackage{natbib}
\usepackage{stmaryrd}
\bibpunct{(}{)}{;}{a}{}{,} % to follow the A&A style

\makeatletter
\def\hlinewd#1{%
\noalign{\ifnum0=`}\fi\hrule \@height #1 %
\futurelet\reserved@a\@xhline}
\makeatother

\newfont{\gwpfont}{cmssq8 scaled 1000}
\newcommand{\rexcess}{{\gwpfont REXCESS}}
\newcommand{\excpres}{{\gwpfont EXCPRES}}

\def\xmm{XMM-{\it Newton} }

\begin{document}
   \title{The dark matter distribution in z$\sim$0.5 clusters
of galaxies}
   \subtitle{I : Determining scaling relations
with weak lensing masses
%}
   \thanks{Based on observations obtained with MegaPrime/MegaCam, a
joint project of CFHT and CEA/DAPNIA, at the Canada-France-Hawaii
Telescope (CFHT) which is operated by the National Research Council
(NRC) of Canada, the Institut National des Science de l'Univers of the
Centre National de la Recherche Scientifique (CNRS) of France, and the
University of Hawaii.}
}

   \author{
          G. Fo\"ex\inst{1,2}
          \and
          G. Soucail\inst{1,2}
          \and
          E. Pointecouteau\inst{1,3}
	  \and
          M. Arnaud\inst{4}
	  \and
          M. Limousin\inst{5,6}
	  \and
          G.W. Pratt\inst{4}
          }
   \institute{
          Universit\'e de Toulouse; UPS-Observatoire Midi-Pyr\'en\'ees; IRAP; Toulouse, France
          \and
          CNRS; Institut de Recherche en Astrophysique et
Plan\'etologie (IRAP); 14 Avenue Edouard Belin, F--31400 Toulouse,  France
         \and
             CNRS; Institut de Recherche en Astrophysique et
Plan\'etologie (IRAP); 9 avenue Colonel Roche, F--31028 Toulouse cedex 4, France
         \and
             Laboratoire AIM; IRFU/Service d'Astrophysique; CEA/DSM; CNRS and 
Universit\'{e} Paris Diderot; B\^{a}t. 709, CEA-Saclay, F-91191 Gif-sur-Yvette Cedex, France
         \and
	      Laboratoire d'Astrophysique de Marseille (LAM); Universit\'e d'Aix-Marseille \& 
CNRS; UMR7326; 38 rue Fr\'ed\'eric Joliot-Curie, F-13388 Marseille Cedex 13, France
         \and
	      Dark Cosmology Centre, Niels Bohr Institute, University of Copenhagen; 
               Juliane Maries Vej 30, DK-2100 Copenhagen, Denmark
             }

   \date{Received ; accepted }

  \abstract  
  % context heading (optional)
  % {} leave it empty if necessary  
  % aims heading (mandatory)
  % methods heading (mandatory)
  % results heading (mandatory)
  % conclusions heading (optional), leave it empty if necessary
   {The total mass of clusters of galaxies is a key parameter to study
massive halos. It relates to numerous gravitational and baryonic
processes at play in the framework of large scale structure formation,
thus rendering its determination important but challenging.  From a
sample of the 11 X-ray bright clusters selected from the \excpres\
sample, we investigate the optical and X-ray properties of clusters with
respect to their total mass  derived from weak gravitational lensing. From
multi-color wide field imaging obtained with MegaCam at CFHT, we derive
the shear profile of each individual cluster of galaxies. We perform
a careful investigation of all systematic sources related to the weak
lensing mass determination. The weak lensing masses are then compared
to the X-ray masses obtained from the analysis of \xmm\ observations and
assuming hydrostatic equilibrium.  We find a good agreement between the
two mass proxies although a few outliers with either perturbed morphology
or poor quality data prevent to derive robust mass estimates. The weak
lensing mass is also correlated with the optical richness and the total
optical luminosity, as well as with the X-ray luminosity, to provide
scaling relations within the redshift range $0.4<z<0.6$.  These relations
are in good agreement with previous works at lower redshifts. For the
$L_X-M$ relation we combine our sample with two other cluster and group
samples from the literature, thus covering two decades in mass and
X-ray luminosity, with a regular and coherent correlation between the
two physical quantities.}


   \maketitle

%
%________________________________________________________________

\section{Introduction}

Clusters of galaxies have been attracting considerable interest
for their cosmological applications since a few decades. The
observed cluster abundance, converted into a mass function,
can be confronted to cosmological numerical simulations to
put tight constraints on cosmological parameters like the
amplitude of matter fluctuations or the dark energy density
\citep{bahcall92,white93,haiman01,wang04,albrecht06,mandelbaum07,rozo09}.
The main limitation in the use of this mass function is the
practical determination of the masses themselves.  In the simplest
model of structure formation and evolution which involves a pure
gravitational collapse of dark matter halos, groups and clusters are
expected to form a population of self-similar objects characterized by
simple relations linking the total mass to other physical quantities
\citep{kaiser86}. These scaling relations represent an efficient way to convert simple observables to masses for
large samples of objects. However, the link between the total mass of
an object and its baryonic tracers is not fully understood. Several
processes break the self-similarity and introduce a scatter around the
theoretical relation that needs to be accounted for when determining
the mass function (see \citealt{voit05} for a review). Hydrodynamic
simulations offers a way to complete this model. They can define more
realistic relations between the cluster total mass and the X-ray
observables (e.g. \citealt{kravtsov06,nagai07}). But the physics of the
baryons within these simulations is not well constrained. Therefore, it
is of prime importance to characterize and calibrate scaling relations
with observational results based on accurate direct mass measurements.
Such empirical relations can also improve our understanding of several
non-gravitational processes disturbing the 
evolution of the large-scale structures.

Among the several ways to derive the total mass of a cluster of galaxies, an efficient one is to use the X-ray emission of the intra-cluster gas. The distribution
in density and temperature can be measured and, assuming
the hydrostatic equilibrium of the cluster, it is possible to
derive the dark matter potential shape. This kind of approach
has been applied successfully on many X-ray observations 
\citep{pointecouteau05,vikhlinin06,buote07}.  However, masses derived
from the X-ray measurements suffer systematics from non-relaxed objects
or from non-thermal pressure support, and tends to underestimate the
true total mass of clusters by $\sim 10-15$\% as seen in numerical
simulations \citep{kay04,nagai07,lau09,meneghetti10} and suggested by observational results (e.g. \citealt{mahdavi08}).

Gravitational lensing is another way to derive the total projected mass,
without any assumption about the dynamical state of the cluster. The strong
lensing regime that occurs in the center of some clusters leads to
elongated arcs and multiple images of background galaxies which probe the
central mass distribution.  In parallel, the weak lensing regime leads
to estimates of the total mass through the statistical distortion of
background galaxies up to large projected distances from the cluster
center (e.g. \citealt{bartelmann01} for a review). However, this method
gives reliable results only for massive clusters in which the weak
lensing signal is detectable with high enough accuracy. It is also
limited to intermediate redshift clusters for which the distant
background galaxies are far enough behind the cluster. Gravitational
lensing masses are also subject to systematics such as the determination
of the redshift distribution of the sources or projection effects
of matter structures near the cluster or along the line-of-sight
\citep{metzler01,hoekstra03,meneghetti10}.

Because all these mass estimators have their own weaknesses and limits and
because they fail to properly recover the total mass in specific cases,
there has been attempts to provide joint analysis of the mass profiles,
combining strong and weak lensing \citep{kneib03,bradac05,merten10,oguri11} or lensing
and X-ray measurements \citep{mahdavi07,morandi1423}. But the difficulty
to get data of similar "quality" to combine them efficiently has up to
now limited the use of such analysis.

In order to better quantify the scaling relations of clusters with
weak lensing masses, we present in this paper the study of a sample
of 11 intermediate redshift clusters which are part of the \excpres\
sample. These clusters have been observed in X-ray with  \xmm\ (Arnaud
et al. in preparation) and in wide field optical imaging with Megacam at
the Canada-France-Hawaii Telescope (\citealt{foex12a}, hereafter Paper 2).
The paper is organized as follows. In section 2 we present the sample
of 11 clusters selected for this study. Section 3 is dedicated to the
lensing methodology with special attention to the shape measurements
of the background galaxies and the normalization of the lensing signal.
Lensing masses are determined from the shear signal in section 4 where
we also make the comparison with the X-ray masses.  Scaling relations
are built and discussed in section 5 and some conclusions are drawn in
section 6. In all the paper, we assume a standard $\Lambda$-CDM cosmology
with $\Omega_{M}=0.3$, $\Omega_{\Lambda}=0.7$ and $H_{0}=70 \ h_{70}\
\mathrm{km\,s^{-1}\,Mpc^{-1}}$.

%__________________________________________________________________

\section{The cluster sample description}

The 11 clusters used in the present study are a sub-sample of the
\excpres\ sample ({\it Evolution of X-ray galaxy Cluster Properties
in a REpresentative Sample}, Arnaud et al. in preparation). This is a
representative X-ray selected sample of clusters in the redshift range
$0.4  < z < 0.6$, compiled from published X-ray selected samples at the
end of 2004 (e.g., EMSS, NORAS, REFLEX, B-SHARC, S-SHARC, 160SD and WARPS-I samples). \excpres\ is designed
to study the evolution of the X-ray properties of clusters. Objects are
selected only on the basis of their X-ray luminosities, as measured in
the aforementioned surveys, and their redshift. The \excpres\ clusters
are selected in the $L_X-z$ plane around a median redshift of $z=0.5$, so that there are about 5 clusters in roughly equal logarithmically-spaced luminosity bins (factor of $\sim 2$ variation between bins). No selection is added with respect to the dynamical state, ensuring that no bias is
introduced by the sample construction process, making it ideal for the
study of scaling relations and their evolution. The final sample contains
20 clusters.\\
The main issue to deal with such a sample of clusters is the selection bias. The 3 clusters from the surveys with a strict X-ray surface brightness limit (B-SHARC and REFLEX) are at 1-2 times the survey limit, i.e. in the regime where the bias is not negligible. For the other clusters from the EMSS and NORAS surveys, that have no defined flux limits, it is difficult to say wether there is a bias. Note however that NORAS clusters are below the $3e^{-12}$ flux limit where the survey is 50\% complete, so likely introducing a bias. Therefore, there is most likely a Malmquist bias, that in principle should be taken into account for the study of the $L_{X}-M$ relation. However it is basically impossible to estimate precisely this bias due to the variety of cluster surveys from which we constructed the \excpres\ sample. Note also that both the $L_{X}-M$ relations from \cite{rykoff08b} and \cite{leauthaud10} (compared and combined with our results in section 5.3) are not corrected for the Malmquist bias. This issue represents a limitation for precise studies of the $L_{X}-M$ scaling relation.\\
However for the $M_{X}-M_{WL}$ mass relation, the original main goal of this paper, the Malmquist bias effect should not be important. It could play a role via possible residual bias in dynamical state: X-ray surveys preferentially select highly peaked cool core (thus relaxed) clusters close to the survey flux limit because the are over-luminous for their mass. Thus the sample may contain more relaxed clusters than the underlying population. This may affect the $M_{X}-M_{WL}$ relation as one expect better agreement for relaxed objects. However in view of the error bars and dispersion (see Figure \ref{fig:X_vs_WL}), this is not the major worry.\\

Note that \excpres\ is deliberately built in a similar way to the local
representative sample \rexcess\ \citep{bohringer07}; so the scaling and
structural properties of \rexcess\ \citep{croston08,pratt09, arnaud10}
will be used as a local reference for \excpres.

Only clusters with an X-ray luminosity $L_X > 5 \times 10^{44}$~erg/s
in the [0.5-2.0]~keV band within the detection radius were selected
for an optical follow-up.  This threshold in $L_X$ ensures that the
total mass of each cluster is high enough to provide significant weak
lensing signal. These measured \xmm\ luminosities were converted into
the [0.1-2.4]~keV energy band for further analysis and comparison with
other works (see Sec.~5.3). They are reported Table~\ref{table:clusters}
together with global temperatures derived within apertures maximizing
the signal-to-noise ratio for the \xmm\ spectroscopic analysis.  In the
framework of the \excpres\ project, the weak lensing measurement of these
11 clusters allows a one-to-one comparison of the X-ray and weak lensing
mass proxies to assess the reliability of the total mass estimate from
X-rays data only.

\begin{table*}
\caption{General properties of the clusters. Columns: (1) Cluster
Name. (2,3) Equatorial coordinates of the X-ray peak. (4) Redshift.
(5) X-ray luminosity in the \xmm\ [0.1-2.4]~keV energy band within the
detection radius, $R_{det}$. (5) Detection radius. (6) Global
spectroscopic \xmm\ temperature, non core excised. (7) Detection of
strong lensing in the CFHT data.}
\label{table:clusters}
\centering 
\begin{tabular}{l c c c c c c c}
\hline\hline\noalign{\smallskip}
Cluster & RA & Dec & z & $L_{X} [0.1,2.4] \mathrm{keV}$ & $R_{det}$ & $kT$
 & strong lens \\
   & (J2000) & (J2000) &  & ($h_{70}^{-2} \ 10^{44} \ \mathrm{erg \,
s}^{-1}$) & (arcmin) & (keV) & (Y/N)\\
\noalign{\smallskip}\hline\noalign{\smallskip}
MS 0015.9+1609 & $00^{h}18^{m}33.26^{s}$ & $+16^{\circ}26{'}12.9{''}$
& 0.541 & 16.3 $\pm$ 0.1 & 5.1 & $9.0\pm 0.2$ & N\\
MS 0451.6--0305 & $04^{h}54^{m}10.85^{s}$ & $-03^{\circ}00{'}57.0{''}$
& 0.537 & 15.2 $\pm$ 0.1 & 3.7 & $8.8 \pm 0.3$ & Y\\
RXC J0856.1+3756 & $08^{h}56^{m}12.69^{s}$ &
$+37^{\circ}56{'}15.0{''}$ & 0.411 & 4.25 $\pm$ 0.1 & 3.8 & $6.9 \pm
0.4$ & N\\
RX J0943.0+4659 & $09^{h}42^{m}56.60^{s}$ & $+46^{\circ}59{'}22.0{''}$
& 0.407 & 4.56 $\pm$ 0.1 & 4.1 & $5.4 \pm 0.2$ & N\\
RXC J1003.0+3254 & $10^{h}03^{m}04.62^{s}$ &
$+32^{\circ}53{'}40.6{''}$ & 0.416 & 2.80 $\pm$ 0.2 & 3.5 & $3.6 \pm
0.2$ & N\\
RX J1120.1+4318 & $11^{h}20^{m}07.47^{s}$ & $+43^{\circ}18{'}06.0{''}$
& 0.612 & 5.82 $\pm$ 0.3 & 2.1 & $5.2 \pm 0.3$ & N\\
RXC J1206.2--0848 & $12^{h}06^{m}12.13^{s}$ &
$-08^{\circ}48{'}03.6{''}$ & 0.441 & 20.7 $\pm$ 0.1 & 6.4 & $9.5 \pm
0.2$ & Y\\
MS 1241.5+1710 & $12^{h}44^{m}01.46^{s}$ & $+16^{\circ}53{'}43.9{''}$
& 0.549 & 6.86 $\pm$ 0.2 & 3.5 & $4.6 \pm 0.1$ & N\\
RX J1347.5--1144 & $13^{h}47^{m}32.00^{s}$ &
$-11^{\circ}45{'}42.0{''}$ & 0.451 & 41.8 $\pm$ 0.1 & 4.9 & $11.4 \pm
0.2$ & Y\\
MS 1621.5+2640 & $16^{h}23^{m}35.16^{s}$ & $+26^{\circ}34{'}28.2{''}$
& 0.426 & 5.09 $\pm$ 0.9 & 4.2 & $5.8 \pm 0.7$ & Y\\
RX J2228.5+2036 & $22^{h}28^{m}33.73^{s}$ & $+20^{\circ}37{'}15.9{''}$
& 0.412 & 13.3 $\pm$ 0.1 & 8.8 & $7.8 \pm 0.3$ & N\\
\noalign{\smallskip}\hline
\end{tabular}
\end{table*}

\section{Weak lensing}
Gravitational lensing distorts the intrinsic image of the background
galaxies.  This is usually quantified by the change in the shape
parameters of the sources. But the intrinsic shape is convolved by the
effect of the atmosphere and the instrumental distortion before it is
measured on CCD images. The PSF correction is therefore a critical step in
the weak lensing analysis and it has to be done with great care. This is
usually validated thanks to realistic simulations of data.  We describe
below the details of how we implemented our weak lensing pipeline up to
the measure of the total mass of the clusters.

\subsection{Objects selection}
The very first step of any weak lensing analysis consists in detecting
and sorting galaxies and stars. This is explained in details in Paper
1, a methodology that is summarized as follows. Using {\sc SExtractor}
in dual mode, we detect the objects on the $\chi_{gri}^{2}$ image while
astrometric and photometric parameters are measured on each individual
$g'$, $r'$, $i'$ and $z'$ images. Stars, galaxies and false detections are
sorted according to several criteria: position in the magnitude/central
flux diagram which displays the {\it  star branch}, size with respect to
the size of the PSF, and stellarity index, according to the CLASS\_STAR
parameter. Objects in masked area are also removed.  After this selection
step, galaxy densities are of the order of $30 \,\mathrm{arcmin^{-2}}$,
and the completeness magnitude ranges from 24.5 to 25 in the $r'$ band,
depending on the data quality in the different cluster fields.

In order to correct for the PSF smearing, we use the {\sc Im2shape}
software \citep{bridle02}, which has already demonstrated its
potential to recover the intrinsic shape of the galaxies
\citep{cypriano04, bardeau05, bardeau07, limousin07a, limousin07b}.
Starting with a given model for the shape of each object, the code
convolves it with the local estimate of the PSF. For simplicity, both
the PSF and the object are modeled with a single elliptical gaussian profile. Exploring the
space parameters with a MCMC sampler, the most likely model is found
by minimizing the residuals, the whole distribution being used to
estimate robust statistical errors on each parameter.

\begin{figure*}
\center
\includegraphics[width=15cm]{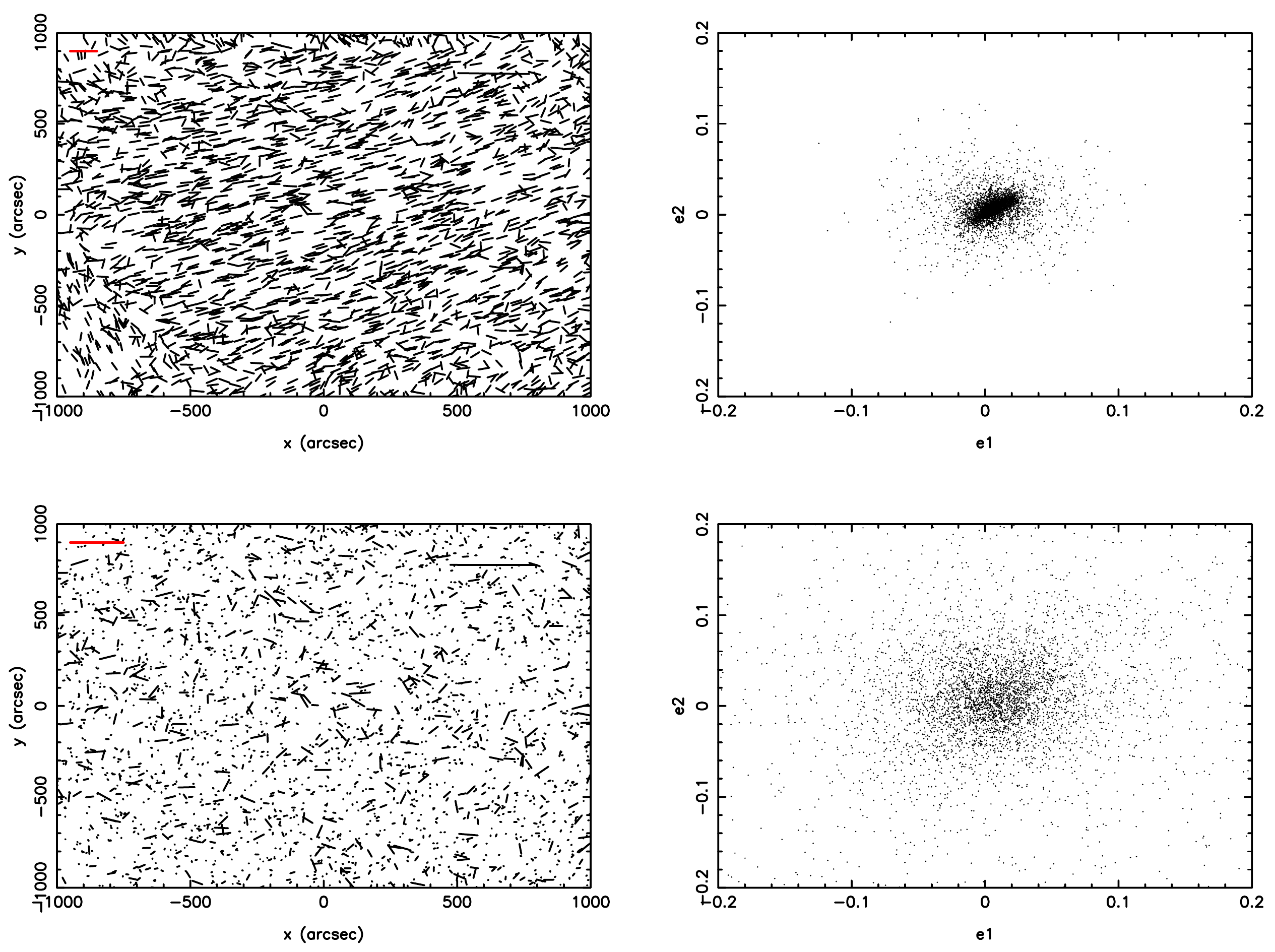}
\caption{PSF treatment applied to stars: on the top-left panel we show the stars ellipticity before the {\sc Im2shape} deconvolution. The corresponding distribution in terms of ellipticity components (e1,e2) is shown in the top-right panel. Bottom-left panel shows the stars after the deconvolution by the PSF field (see text) and the bottom-right panel the corresponding distribution of (e1,e2), which is more uniformly distributed than before the deconvolution. In both left panels is shown in red the scale corrsponding to a semi-major axis of 1''.}
\label{fig:PSFmap} 
\end{figure*}

\begin{figure*}
\center
\rotatebox{0}{
\includegraphics[width=15cm]{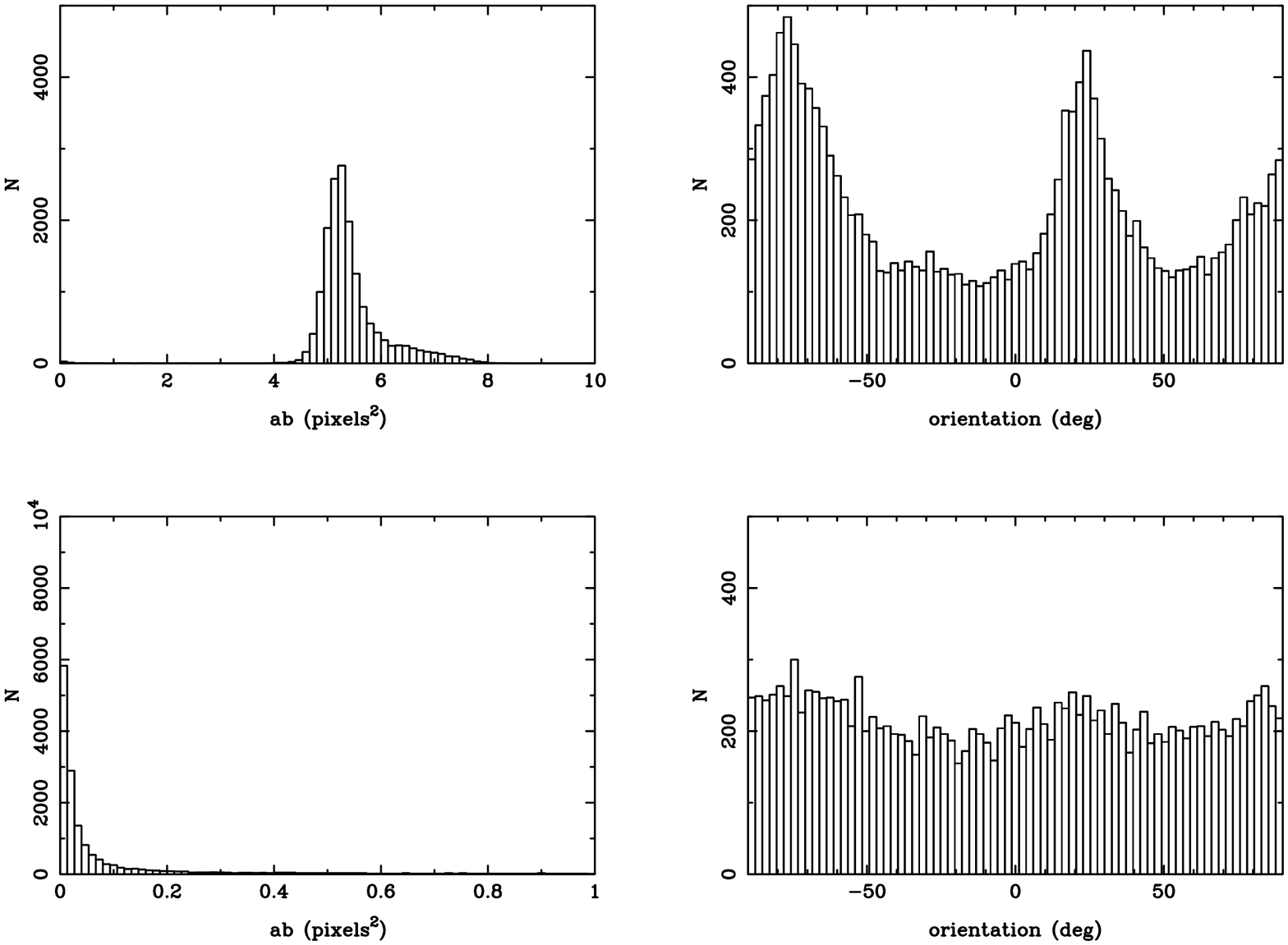}
}
\caption{Distribution of the area (left) and orientation (right) of
the  stars selected in the field of RX J2228.5+2036. The top panels
represent the stars before the PSF 
deconvolution by {\sc Im2shape} while the bottom panels show the
distribution after deconvolution, with an average size close to 
point\_like objects and a uniform orientation of the stars.
} 
\label{fig:im2_2228} 
\end{figure*}

We first determine the local PSF by looking at the shape of the stars. The resulting PSF field is cleaned and smoothed by looking at the 10 nearest stars at each point and removing those that differ by more than $1.5\sigma$ from the local average shape. This ensures to avoid any jump in the PSF pattern, e.g. near masked area. We also exclude objects having an ellipticity larger than 0.2, thus reducing the contamination of the stars catalog by false detections and faint galaxies. The PSF map over the whole field of view is then obtained by averaging the ellipticities of the 5 nearest
stars at each galaxy position. We checked
that our {\sc Im2shape} implementation can recover point-like objects
by applying this PSF correction to each star. Figure \ref{fig:im2_2228}
shows the results of the shape parameters measurements for these stars:
the size distribution is dominated by point-sources and the orientation
is more uniformly distributed after the PSF correction. It is also shown Figure \ref{fig:PSF} where the regular pattern of the stars ellipticities due to the PSF (top row) disappear after the deconvolution.

\subsection{Validation of the shape measurements with simulated data}
To check the performance of our {\sc Im2shape} implementation,
we tested our pipeline on the simulated fields provided by the STEP
collaboration (the Shear Testing Programme, \citealt{heymans06}). STEP1
simulations were built in order to compare the accuracy of different
lensing pipelines for ground-based images. They provide a set of images
reproducing CFHT-like observations with representative densities of stars and galaxies. These simulations combine several PSF models and shear intensities. We ran our lensing pipeline on each corresponding data set, compare our results to the input configurations and derive an average shear calibration bias and PSF residual. 

%For each data set, the mean shear $\gamma_{1,2}$ is computed and
%averaged over the 64 images. Then for each PSF model the 5 shear
%values are fitted with the parameters $m$, $q$ and $c$ defined as:
%\begin{eqnarray}
%& & \langle\gamma_{1}\rangle-\gamma_{1}^{true} =  q \ 
%(\gamma_{1}^{true})^2+m \ \gamma_{1}^{true}+c_{1}\\
%& & \nonumber c_{2} = \langle\gamma_{2}\rangle
%\end{eqnarray}
%A linear response of the method to the shear is characterized by $q=0$,
%which is valid in our case.  In the absence of shot noise and PSF
%systematics, we expect to have $c_{1}=0$ and $c_{2}=0$ since the input
%external shear is defined by $\gamma_{2}^{true}=0$. $m$ represents the
%calibration bias and should be as close as possible to 0. The
%non-linearity parameter $\langle q\rangle$, the calibration bias
%$\langle m\rangle$, and the PSF residuals
%$\sigma_{c}=\sqrt{\sigma_{c1}^{2}+\sigma_{c2}^{2}}$ are finally
%obtained by averaging the values of the fits over the 6 models of
%PSF.

%\begin{figure}
%\center
%\includegraphics[width=\hsize]{step2}
%\caption{Calibration bias $\langle m\rangle$ and PSF residuals
%$\sigma_{c}$ measured for different lensing methods with the STEP1
%simulations \citep{heymans06}. The shaded
%region corresponds to a calibration bias smaller than 7\%. Our implementation
%of {\sc  Im2shape} is labeled GF, to be compared with the more
%complex implementation of {\sc Im2shape} proposed by S. Bridle (SB).}
%\label{fig:step1} 
%\end{figure}

Our implementation of {\sc Im2shape} differs slightly from the one
proposed by S. Bridle in the STEP1 simulations: for simplicity
and efficiency we fitted both the PSF and the galaxy shapes by a single
elliptical gaussian instead of 2 concentric ones. This avoids to deal with 2 different values of the galaxies ellipticity when measuring the shear profiles.
%As we can see in Figure
%\ref{fig:step1}, 
Despite this simplification, our method is very satisfactory as we obtain a calibration bias $\langle
m\rangle\simeq-0.1 \pm 0.02$ and a PSF residual $\sigma_{c} \simeq 2.10^{-3}$, a
value consistent with shot noise (see eq. 11 of  \citealt{heymans06}). These two values can be compared to the results from other weak lensing methods that are summarized on Figure 3 of \cite{heymans06}. So in the present case we introduce an underestimation of the true shear around 10\%
but with no significant systematics. In the rest of the paper we will
therefore increase all measures of the shear by a 10\% factor before we use them to derive cluster masses.
% This leads to a 10 to 15\% increase in the final masses. 

\subsection{Shear radial profiles}
To quantify the shape of a galaxy, we use the complex ellipticity
defined by \citet{bonnet95}. It relates the tensor of the galaxy
brightness second moments to its shape. Applying the lensing
transformation between the source and image planes, one find the
simple relation \citep{seitz1997}:
\begin{equation}
e^{(s)} = \left\{ \begin{array}{c}
\dfrac{e - g}{1-g^*e} \quad {\rm for} \: |g| \leq 1 \\
\noalign{\smallskip}\\
\dfrac{1 - g e^*}{e^*-g^*}  \quad {\rm for} \: |g|>1
\end{array}
\right.
\end{equation}
where $*$ denotes the complex conjugate, $e^{(s)}$ being the complex
ellipticity in the source plan and $e$ in the image plan. $g$ is the
reduced shear
\begin{equation}
g=\frac{\gamma} {1-\kappa}
\end{equation}
which is a non-linear function of the two lensing functions: the
(complex) shear $\gamma$ and the convergence $\kappa$ which is related
to the projected mass density.

We assume spherical symmetry of the mass distribution and
we consider that unlensed galaxies are randomly oriented on
the sky plane. This is observationally confirmed since the
distribution of the field galaxy ellipticities can be fitted
by a gaussian $\mathcal{N}(0,\sigma_{e}^{2}\approx0.25^{2})$
\citep{tyson88,brainerd96}. In the weak lensing approximation (i.e.
$\kappa \ll 1$) we get an unbiased estimator of the reduced shear
by averaging the shape of background galaxies in concentric annuli
around the cluster center. Spherical symmetry also implies that
the averaged radial component of the shear vanishes, i.e. $\langle
e_{\perp}\rangle=0$, while the tangential component of the lensed
galaxies traces the reduced shear $\langle e_{\sslash}\rangle=g$. We also
assume that $\langle e^{(s)}\rangle=0$ with an intrinsic dispersion
$\sigma_{e}\approx0.25$. The measure of this averaged tangential
component $\langle e_{\sslash}\rangle$ at different distances provides
a reduced shear profile $g(r)$ which is used to recover the total mass
of the deflector.

\begin{figure}
\center
\includegraphics[width=\hsize]{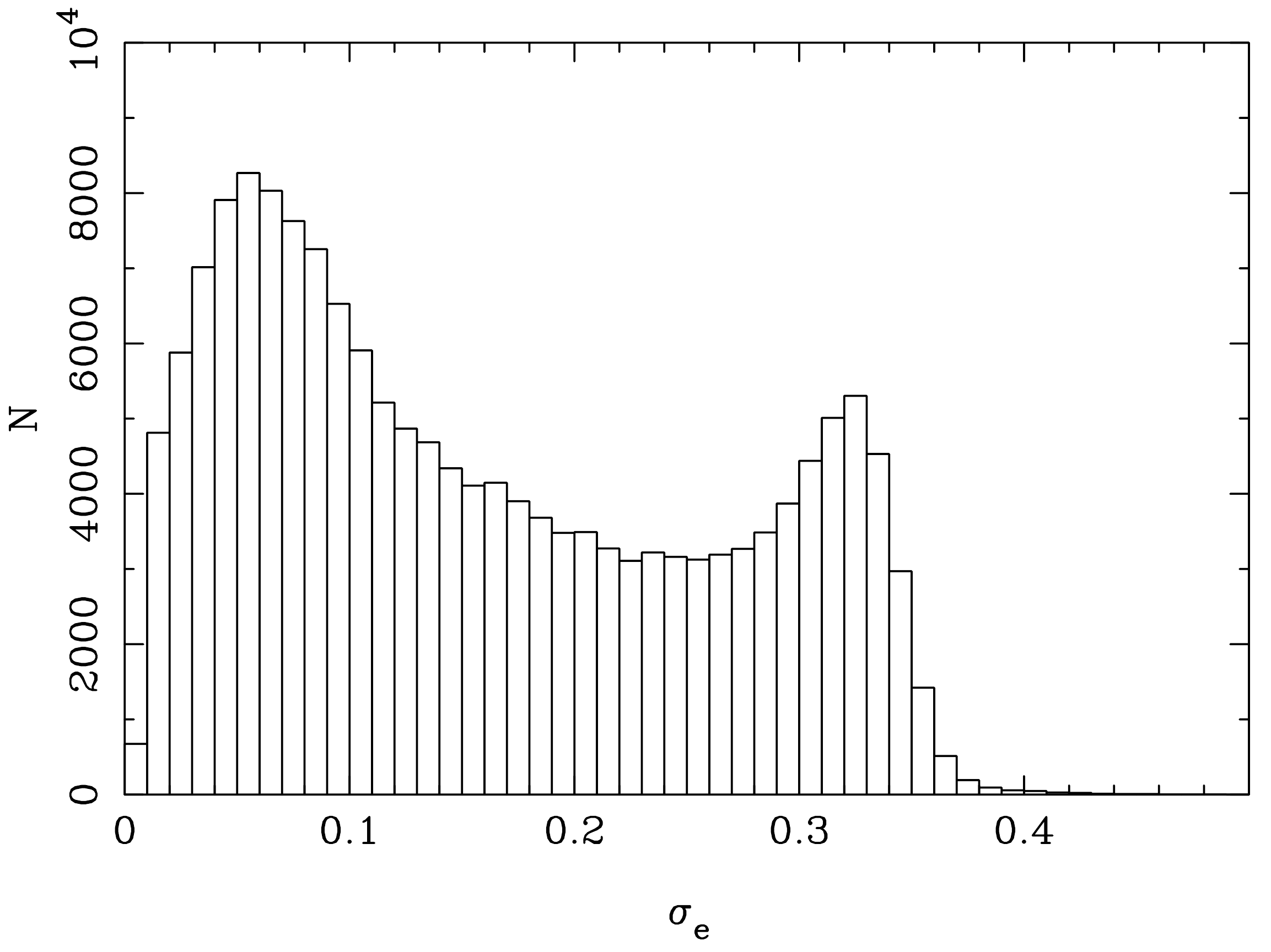}
\caption{Distribution of the errors on the galaxies
ellipticity estimator given by {\sc Im2shape} (here for RX
J2228.5+2036). The minimum of the histogram is reached for an
uncertainty $\sigma_{e}\sim0.25$. This value gives a natural 
cut to clean up the catalogs
of galaxies used to measure the shear profiles.}
\label{fig:err} 
\end{figure}

We fix the center of the lens at the position of the brightest
galaxy (BCG). The non-parametric mass reconstructions of the cluster
sample that will be presented in Paper 2, show separations between the cluster mass
center and the location of the BCG smaller than 30\arcsec\ for all
clusters except the most disturbed ones or those with a low S/N mass
reconstruction. Real shifts between the BCG and the mass center are expected, especially in the most massive dark matter haloes or in non-relaxed clusters (see e.g. \citealt{skibba10} for a discussion of the central galaxy paradigm). However, \citet{dietrich11} also showed with simulated data that
for mass reconstructions done with ground-based data, shifts are observed
with a median value of 1\arcmin\ between the true and the reconstructed
mass center. So here we can not reliably disentangle artefatcs in the 2D mass reconstructions from real shifts between the BCG and the mass center. Moreover, given the redshifts and mass of the clusters, small miss-centerings disturb only
the central data points in the shear profile. As discussed in \cite{mandelbaum10}, excluding the central parts of the profiles in the fitting procedure reduces the underestimation of the mass because of miss-centering. Basic simulations (lensing a typical catalog of sources by a cluster representative of the sample in terms of mass and redshift) show that our fitting methodology gives consistent masses within their error bars up top miss-centering of 1\arcmin. Therefore we consider that in our case the BCG is a good estimator of
the actual mass center and that this assumption does not have any consequence on the total mass determination.

Shear profiles are built in non-overlapping logarithmic annuli with
$r_{out}=1.2r_{in}$, in order to have similar S/N ratios in
each bin. Thanks to the wide field images taken with Megacam, we are able
to measure shear profiles up to 600 to 800\arcsec\
from the BCG. We limit the fit of the shear profile in the central region
to an inner radius set at 25\arcsec\ (which roughly corresponds to the Einstein radius of the clusters) except for 4 clusters presenting a shear signal that is too disturbed
in the central part. In these cases, the limit is fixed to the radius
where the signal becomes significantly positive. This inner limit
reduces the impact of miss-centering and avoids complications due to the
non-linearity of the reduced shear with mass density. It also avoids
the noise induced by galaxies which shape is badly estimated because
they are too close to bright cluster galaxies (e.g.  \citealt{cypriano04}).

Within these limits, 10 to 15 data points are available in the shear
profile. In order to improve the statistical weight of the results, we
have chosen to construct 25 {\it shifted} profiles by moving the
position of the lower bound of the first bin by 1/25 of its width.
This allows to reduce sampling effects in the first points of the
profile where the averaged tangential ellipticity can significantly
change according to the way the galaxies are binned. Each of these 25
profiles are fitted using a standard $\chi^{2}$ minimization {\bf (the distribution of field galaxies ellipticity being gaussian, e.g. \cite{brainerd96}, the likelihood follows gaussian statistics)}. Doing so, we limit the correlation that would arise with overlapping bins.

%The best
%fit parameters of the model then correspond to:
%\begin{equation}
%\boldsymbol{\theta}_{best}=\min\limits_{\boldsymbol{\theta}} 
%\left[\sum_{i=1}^{N}\frac{\left(g(\boldsymbol{\theta},r_{i})-g_{i}\right)^{2}}
%{\sigma_{i}^{2}}\right]
%\end{equation}
%where $N$ is the number of points in the profile, $r_{i}$ the distance
%to the center of the $i-th$ bin, $\boldsymbol{\theta}$ represents the
%free parameters of the analytical model and $g_{i}$ and $\sigma_{i}$
%are the signal and error measured in the $i-th$ bin. Note that we
%include at this level the calibration factor found with the STEP1
%simulations to correct the value of the shear signal.

To improve the
shear estimator, we weighted the average value of the
tangential component of the galaxies ellipticity with the inverse of
the ellipticities distribution variance
$\omega_{j}=1/\sigma_{\sslash,j}^{2}$:
\begin{equation}
g_{i}=\frac{\sum_{j=0}^{N_{i}}\omega_{j}e_{\sslash}^{j}} 
{\sum_{j=0}^{N_{i}}\omega_{j}}.
\end{equation}
$\sigma_{\sslash}$ is the quadratic sum of the errors
$(\sigma_{e1},\sigma_{e2})$ given by {\sc Im2shape} for each object and 
the intrinsic dispersion of the galaxies ellipticities
$\sigma_{int}=0.25$. Note that we also cut our catalogs of galaxies by
limiting the uncertainties returned by {\sc Im2shape} to
$(\sigma_{e1},\sigma_{e2})<0.25$. This value corresponds to the
average minimum in the uncertainties distributions (Figure
\ref{fig:err}), the objects above this threshold being most likely false detections or very faint/small galaxies. Note also that thanks to our weighting scheme their removal does not have any consequences on the final mass estimates (see \citealt{cypriano04} for details).
% Finally the error that enters the $\chi^{2}$
%minimization is:
%\begin{equation}
%\sigma_{i}=\frac{1}{\sqrt{\sum_{j=0}^{N_{i}}(1/\sigma_{\sslash,j}^{2})}}.
%\end{equation}

To assess the uncertainties on the best fit parameters, we proceed as
follow. Assuming that the measures $g_{i}$ have a gaussian uncertainty
$\sigma_{i}$ (although the distribution of the errors appears to be highly non-gaussian on Figure \ref{fig:err}, ellipticities for each galaxy generated by {\sc Im2shape} have a gaussian distribution), we generated 1000 Monte Carlo draws of each of the 25
shifted shear profiles, and applied the $\chi^{2}$ procedure. We then obtain 25000 values of the best fit parameters. The mode of the resulting distribution gives the most likely solution along with the 1, 2 and 3 $\sigma$ asymetric errors. 

\subsection{Normalization of the shear profiles}
\begin{figure}
\center
\includegraphics[width=\hsize]{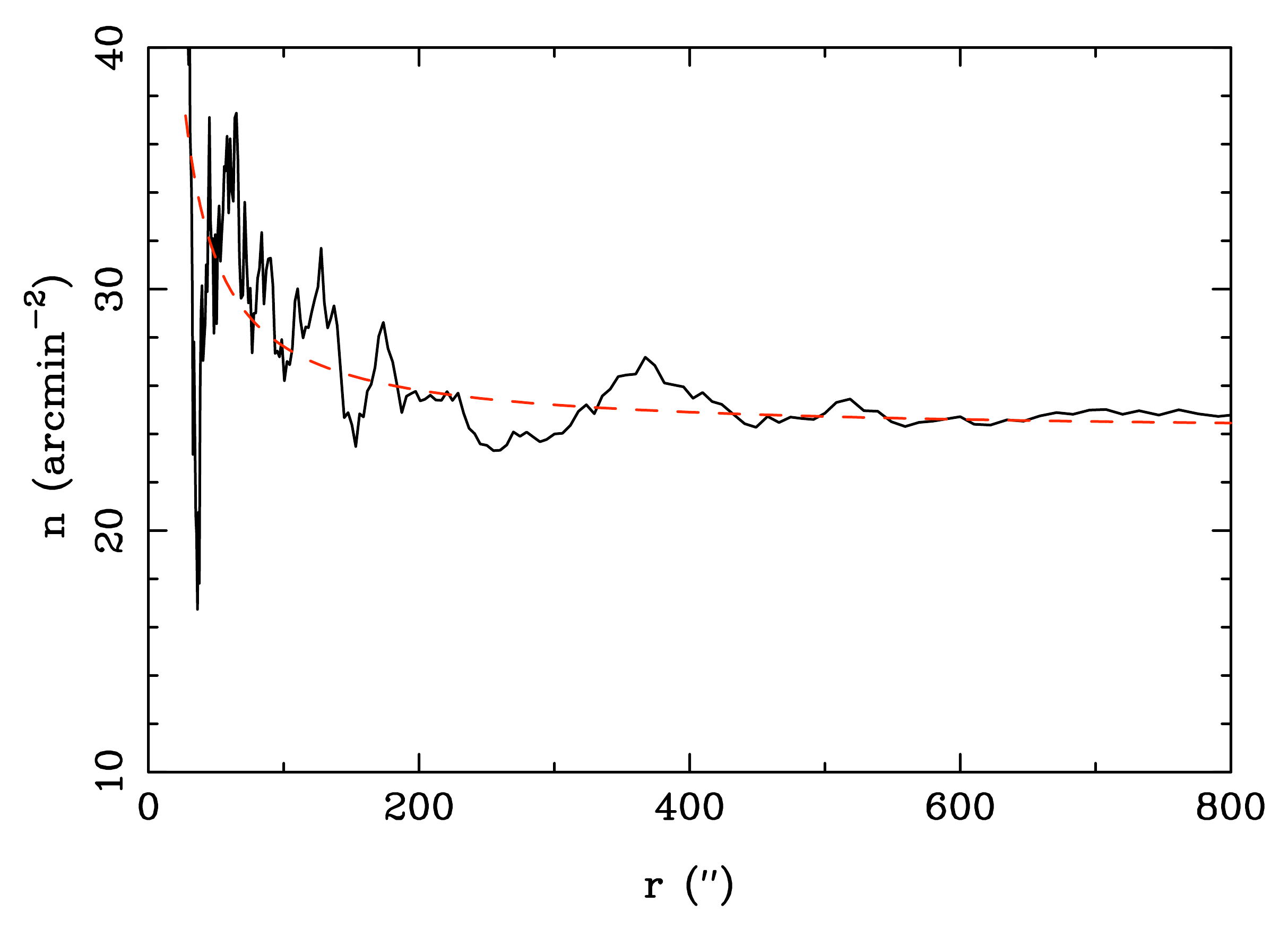}
\caption{Density profile of the background galaxies used for the fit
of the shear profile in the field of MS 1621.5+2640. Despite the
removal of the galaxies within the red sequence, 
a substantial fraction of cluster members remains. The density profile
is fit (dashed red line) to give the boosting factor applied to the
shear profile. 
}
\label{fig:radprof} 
\end{figure}

To limit the contamination of the sources catalogs we introduce several magnitude and color cuts which are discussed in details in Paper 2. In summary, we limit the magnitude range to 21--m\_{comp}+0.5 in $r'$ and remove all objects located within a color-band defined by the bright ellipticals red sequence up to $m_r = 23$. This reduces the galaxy density by about 1/3 even if some contamination by bluer galaxies remains. To overcome
this problem, we followed the method proposed by \cite{hoekstra07}
which assumes that the number density of cluster members simply decreases with
radius as $1/r$.  We adjust the normalization by fitting the profile
of galaxies excess relative to the background level and we correct
the measured shear according to the distance to the cluster center
(Figure~\ref{fig:radprof}). As the richness of a cluster scales with its mass, this contamination is higher for more massive objects. So we apply the renormalization procedure for each cluster instead of using an average correction as done in \cite{hoekstra07}.

Photometric redshift information is also necessary to correct for two
effects which change the normalization of the shear profiles. The first
one is an assessment of the fraction of foreground galaxies in the
catalogs and the second one is the determination of the average redshift
distribution of the background sources. To get a reliable faint galaxies
redshift distribution, we used the photometric redshifts of galaxies
in reference fields, neglecting the cosmic variance between different
fields \citep{cypriano04,hoekstra07,sl2s,oguri09}. Observations in
the CFHTLS Deep fields obtained with the same instrument form the ideal data set for our study \citep{gavazzi07}. They are much deeper than the present observations so we can consider
that their catalogs are complete up to our magnitude limit. We used
the photometric redshifts determined with the {\sc HyperZ} software
\citep{bolzonella00}. They were calibrated and validated by comparison with
spectroscopic redshifts obtained in the VVDS D1 field \citep{lefevre05}
and in the Growth Deep survey \citep{weiner05}. Once our photometric
selection criteria were applied on the CFHTLS catalogs, we
obtained a redshift distribution that is supposed to be representative of the catalogs we use for the weak lensing analysis.

We use this distribution of photometric redshifts to determine the strength of the shear signal via the geometrical
factor $\beta(z_{L},z_{S})=D_{LS}/D_{OS}$ ($D_{LS}$ is the
angular-diameter distance between the lens and the source and $D_{OS}$
the distance to the source). Both the convergence $\kappa$ and the
shear $\gamma$ can be re-written as \citep{hoekstra00}:
\begin{equation}
g=\frac{\gamma}{1-\kappa}=\frac{\beta_{s}\gamma^{\infty}}
{1-\beta_{s}\kappa^{\infty}}
\end{equation}
with $\beta_{s}=\beta/\beta^{\infty}$ and $\beta^{\infty},
\gamma^{\infty}, \kappa^{\infty}$ are the corresponding functions 
for a source at infinity. In the weak lensing approximation,
$g\approx\gamma$, the redshift dependence of the signal is easy to
take into account as the average reduced shear simply equals its value taken with the average geometrical factor.\\
For galaxy clusters at low
redshift, it is straightforward to replace $\langle\beta\rangle$ by the
$\beta$ value of an average redshift $z_{s}$, usually set to 1
\citep{okabe08,radovich08}. In the present case, the geometrical
factor varies significantly with the sources redshifts so we computed the factor $\langle\beta(z)\rangle$ for the whole distribution of field galaxies. We account for the contamination by foreground galaxies given our selection criteria by setting $\beta(z_{phot}<z_{cluster})=0$ (values in a range of 20\% to 30\% depending on the cluster). Note that such an approach is only valid in the weak lensing regime because of the non-linearity of the reduced shear with $\beta$ (see e.g. \citealt{seitz1997,hoekstra00}). However, as with remove the central parts from the fits, we expect these $\langle\beta(z)\rangle$ to be a good approximation. The averaged geometrical factors are given in Table
\ref{table:results1} along with the effective redshift defined as
$\beta(z_{eff})=\langle\beta(z)\rangle$. 

%Although this approach is more rigorous, it still suffers from the
%fact that the reduced shear $g$ is not proportional to $\beta$ and
%that $\langle
%g\rangle\neq g(\langle\beta\rangle)$. An estimate of the error due to
%this approximation is given at first order by (e.g.
%\citealt{seitz1997,hoekstra00}):
%\begin{equation}
%\frac{g(\langle \beta \rangle)}{\langle g(\beta)\rangle }=
%1+\langle \kappa \rangle(1-f)<1
%\end{equation}
%This ratio depends on the fraction $f=\langle \beta
%^{2}\rangle/\langle \beta \rangle^{2}$ and the convergence $\langle
%\kappa \rangle=\langle \beta_{s} \rangle\kappa^{\infty}$. This
%non-linearity effect mainly affects the central part of the shear
%profile and results in an over-estimation of the cluster mass if not
%corrected. The fraction $f$ is obtained with the CFHTLS Deep field
%catalogs of photometric redshifts and is of the order of $1.4-1.9$
%(depending on the cluster redshift). So we applied this correction
%$1+\langle \kappa \rangle(1-f)$ to the analytical models $g(\langle
%\beta \rangle)$ before we fit the measured reduced shear $\langle
%g(\beta)\rangle$.  

\begin{table*}
\caption{Summary of the weak lensing analysis. Columns are (1) cluster name; (2-3) NFW masses; (4) SIS velocity dispersion; (5) Einstein radius for a source at $z_{s}=2$; (6-7) radius corresponding to the NFW masses; (8) average geometrical factor corrected from dilution (see text); (9) effective redshift derived from the average geometrical factor.}
\label{table:results1}
\centering 
\begin{tabular}{l c c c c c c c c}
\hline\hline\noalign{\smallskip}
\multirow{2}{*}{Cluster} & $M_{200}$ & $M_{500}$ & $\sigma_{v}$ & $\theta_{E}$ & $R_{200}$ & $R_{500}$ & $\langle\beta\rangle$ & $z_{eff}$\\
 & ($10^{15}h_{70}^{-1}M_{\odot}$) & ($10^{15}h_{70}^{-1}M_{\odot}$) & ($\mathrm{km.sec^{-1}}$) & ('') & ($h_{70}^{-1}$ kpc) & ($h_{70}^{-1}$ kpc) & &\\
\noalign{\smallskip}\hline\noalign{\smallskip}
MS 0015.9+1609 & $2.56_{-0.41}^{+0.49}$ & $1.77_{-0.29}^{+0.34}$ & $1490_{ -90}^{+  90}$ & $39_{-5}^{+ 5}$ & $ 2330_{-130}^{+ 130}$ & $1520_{ -90}^{+  90}$ & 0.26 & 0.78\\
\noalign{\smallskip}
MS 0451.6--0305& $1.44_{-0.26}^{+0.30}$ & $1.00_{-0.18}^{+0.21}$ & $1280_{ -90}^{+  70}$ & $29_{-4}^{+ 3}$ & $ 1920_{-120}^{+ 120}$ & $1250_{ -80}^{+  80}$ & 0.27 & 0.79\\
\noalign{\smallskip}
RXC J0856.1+3756 & $0.78_{-0.15}^{+0.12}$ & $0.54_{-0.11}^{+0.09}$ & $1000_{ -70}^{+  50}$ & $20_{-3}^{+ 2}$ & $ 1650_{-120}^{+  80}$ & $1070_{ -70}^{+  50}$ & 0.39 & 0.73\\
\noalign{\smallskip}
RX J0943.0+4659 & $0.93_{-0.21}^{+0.18}$ & $0.64_{-0.15}^{+0.12}$ & $1090_{ -100}^{+  60}$ & $24_{-4}^{+ 3}$ & $ 1770_{-160}^{+  90}$ & $1150_{-100}^{+  60}$ & 0.38 & 0.72\\
\noalign{\smallskip}
RXC J1003.0+3254 & $0.74_{-0.13}^{+0.18}$ & $0.51_{-0.09}^{+0.13}$ & $ 990_{ -70}^{+  70}$ & $19_{-3}^{+ 3}$ & $ 1690_{-140}^{+  90}$ & $1100_{ -90}^{+  60}$ & 0.37 & 0.72\\
\noalign{\smallskip}
RX J1120.1+4318 & $0.59_{-0.15}^{+0.26}$ & $0.41_{-0.10}^{+0.18}$ & $1020_{-110}^{+  90}$ & $ 17_{-4}^{+ 3}$ & $ 1460_{-150}^{+ 130}$ & $ 950_{-100}^{+  90}$ & 0.23 & 0.85\\
\noalign{\smallskip}
RXC J1206.2--0848 & $1.51_{-0.18}^{+0.28}$ & $1.05_{-0.12}^{+0.19}$ & $1290_{ -60}^{+  50}$ & $32_{-3}^{+ 3}$ & $ 2030_{ -90}^{+ 110}$ & $1320_{ -60}^{+  70}$ & 0.36 & 0.75\\
\noalign{\smallskip}
MS 1241.5+1710 & $1.39_{-0.30}^{+0.30}$ & $0.96_{-0.21}^{+0.20}$ & $1230_{ -90}^{+ 100}$ & $26_{-4}^{+ 4}$ & $ 1880_{-140}^{+ 130}$ & $1230_{ -90}^{+  80}$ & 0.27 & 0.81\\
\noalign{\smallskip}
RX J1347.5--1144 & $2.56_{-0.31}^{+0.32}$ & $1.77_{-0.21}^{+0.22}$ & $1500_{ -60}^{+  60}$ & $43_{-3}^{+ 3}$ & $ 2400_{-100}^{+  100}$ & $1560_{ -60}^{+  60}$ & 0.35 & 0.76\\
\noalign{\smallskip}
MS 1621.5+2640 & $1.23_{-0.22}^{+0.20}$ & $0.85_{-0.15}^{+0.14}$ & $1200_{ -80}^{+  60}$ & $28_{-4}^{+ 3}$ & $ 1900_{-120}^{+  100}$ & $1240_{ -80}^{+  60}$ & 0.38 & 0.75\\
\noalign{\smallskip}
RX J2228.5+2036 & $0.85_{-0.18}^{+0.21}$ & $0.59_{-0.12}^{+0.14}$ & $1050_{ -80}^{+  80}$ & $22_{-3}^{+ 3}$ & $ 1680_{-120}^{+ 130}$ & $1100_{ -80}^{+  80}$ & 0.39 & 0.75\\
\noalign{\smallskip}\hline
\end{tabular}
\end{table*}

\section{Weak lensing masses}

\subsection{Mass models}
We restrict our analysis to the two mostly used parametric mass models:
the singular isothermal sphere (SIS hereafter) and the NFW model
\citep{navarro95}. The SIS mass model is the most simple one to describe
a relaxed massive sphere with a constant and isotropic velocity dispersion
$\sigma_{v}$. The mass density profile writes as
\[ \rho(r) = \frac{\sigma_{v}^{2}}{2 \pi G \ r^{2}} \]  
and simple expressions are derived for the lensing functions
$\gamma(r)=\kappa(r)=R_{E}/2r$ where the Einstein radius scales as
$R_{E}\propto\beta\sigma_{v}^{2}$. This mass model is a poor description of the actual mass distribution of clusters. So the fitting results of this mass model will not be used to derive the total clusters mass in the rest of this work. However it allows easy comparisons
with other mass estimators such the dynamical analysis of the
galaxies members that are also characterized by a velocity dispersion $\sigma_{v}$. It also provide an estimate of the Einstein radius $R_{E}$ which can be compared to results from strong lensing analysis. This latter quantity requires an estimate of the geometrical factor to be converted from $\sigma_{v}$. Values given in Table \ref{table:results1} are derived using a source redshift $z_{s}=2$ which leads to radii in the range $20''-40''$ (note that these values are not directly estimated from the signal in the central part as done in strong lensing analysis but derived from the fit of the weak lensing shear profile).

The NFW mass profile \citep{navarro97,navarro04} was introduced to model 
the mass distribution of a cold dark matter halo formed in
a gravitational scenario of structures formation. It involves a shape parameter, the concentration $c$,
and a normalization parameter often expressed as the scale radius $r_{s}$:
\[
\rho(r)=\frac{\rho_{c} \ \delta_{c}}{(r/r_{s})(1+r/r_{s})^{2}}
\]
$\rho_{c}$ is the critical mass density of the Universe at the
redshift of the cluster and $\delta_{c}$ is a 
dimensionless factor which is related to the contrast density
of a virialized dark matter halo $\Delta = \overline{\rho}/\rho_{c}
=200 $ by 
\[ 
\delta_c = \frac{200}{3} \ \frac{c^3}{\ln (1+c) - \frac{c}{1+c}}
\]
 We also define $R_{200}=c \ r_{s}$, often called
the virial radius, and the mass enclosed within this radius
\[ 
M_{200}=\frac{800 \ \pi}{3} \ \rho_{c} \ R_{200}^{3}
\] 
$\Delta=200$ roughly corresponds to the virialized part of clusters.
X-ray studies more often make use of $\Delta =500 $, a density contrast
more easily reached by  the X-ray detection. Given the above relations,
it is straightforward to analytically switch masses and radii from one $\Delta$
value to another.

\begin{figure}
\centering
\rotatebox{0}{
\includegraphics[width=0.98\hsize]{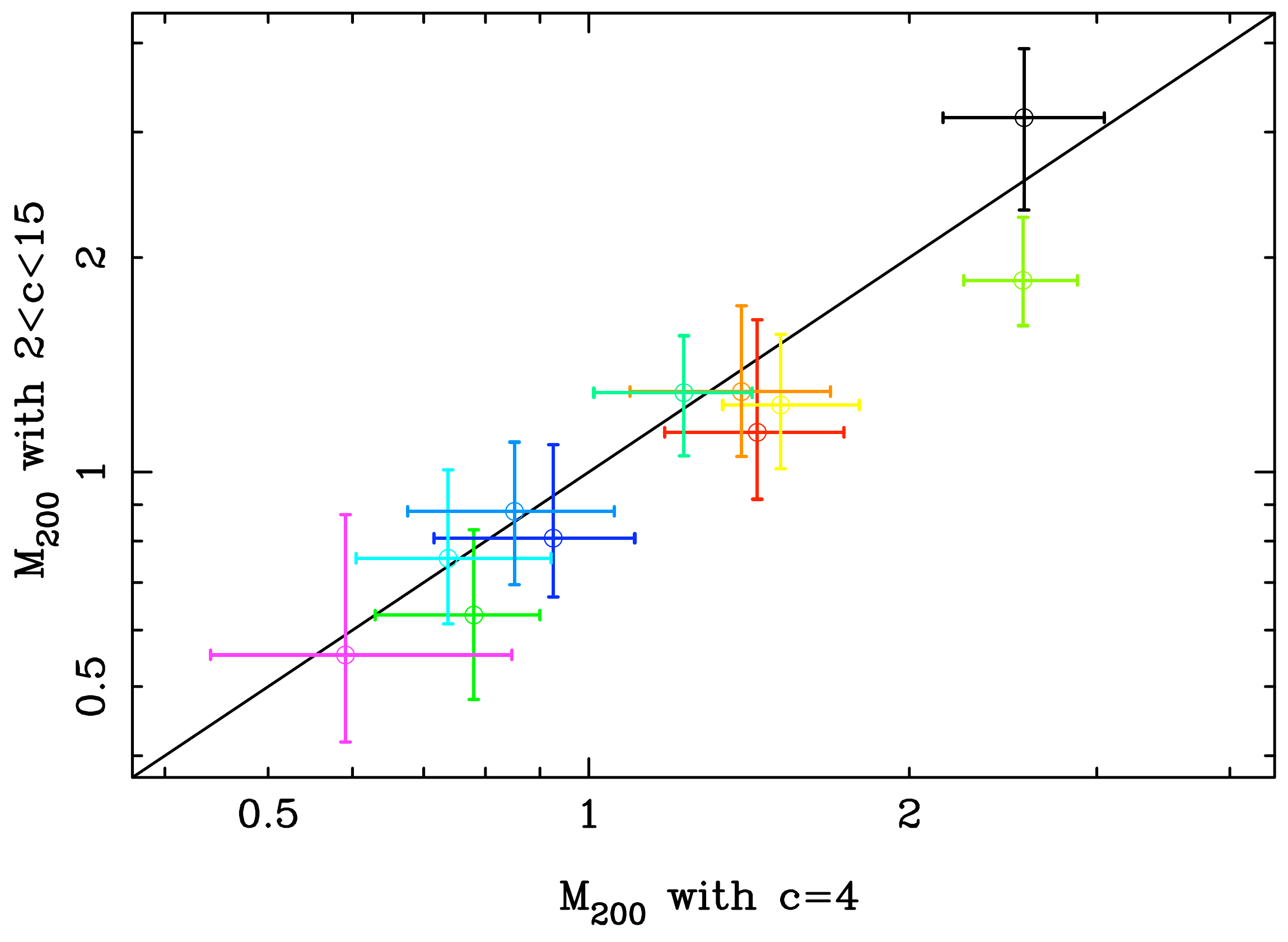}
}
\caption{Comparison between the total mass $M_{200}$ (expressed in
units of $10^{15}\mathrm{h_{70}^{-1}} M_{\odot}$) obtained by fitting
the shear profile with a NFW model. In the first case the
concentration parameter is fixed to $c=4$ while in the second case it
is left as a free parameter varying over the range $2<c<15$,
increasing the mass uncertainties by 25\%\ typically. 
} 
\label{fig:c4_vs_c215} 
\end{figure}

With two free parameters, the NFW model provides more freedom to adjust
the mass profile of galaxy clusters. However, there is a well-known degeneracy
between the total mass $M_{200}$ and the concentration parameter $c$ when
fitting the shear profile in the weak lensing regime. This comes from a lack of information on the
mass distribution near the cluster center. The observed $c-M$ degeneracy
shows a much steeper slope than the theoretical one \citep{okabe10a} and
only a combination of strong and weak lensing can raise it and provide
useful constraints on the concentration parameter. In the present case,
with no strong lensing modeling of all the clusters in the sample,
we decided to fix the concentration parameter and only fit the total mass $M_{200}$. Indeed for most of the
clusters, the fit of both parameters leads to unrealistic values of $c$, either larger than 20 or, in most cases, sticked to the lower limit of the fit $c=2$. Since a low concentration is associated to a lower mass when fitting a given shear profile, this leads to the apparent bias observed Figure \ref{fig:c4_vs_c215}, which is only an artefact due to poor constraints in the central parts of the shear profiles. For a more secure mass estimator we therefore fixed the concentration parameter $c$ to a canonical value for massive clusters,  $c_{200}=4$, and fit the mass profile with only one free parameter.

It is interesting to note that the ratio between the NFW mass $M_{200}$
and the SIS mass integrated within $R_{200}$ is close to one for the whole
sample: $\langle M^{NFW}_{200}/M^{SIS}(R_{200})\rangle=0.98\pm0.06$. This
reflects the fact that in this study of high redshifts clusters of galaxies, weak lensing data only provide constraints on the
total mass at large scales. It cannot efficiently constrain the radial
mass profile contrary to clusters at lower redshifts where some disagreements are found between the 2 mass models \citep{okabe10a}.\\
 In summary, we have in hands for each cluster the
results of the fit by a SIS profile ($\sigma_{v}$ and $R_{E}$) as well
as those by a NFW profile ($M_{200}$ and $R_{200}$). The mass and the
radius taken at $\Delta = {500}$ are also computed for comparison with
the results of the X-ray analysis. All these values are given in Table
\ref{table:results1}. Note that because we are interested here in the calibration of scalings relation and the scatter of individual clusters around the best fits, we do not have included the systematic uncertainty due to projection effects of the large scale structures. This effect should be less than $10\%$ for rich clusters at intermediate redshifts \citep{hoekstra01}. We also did not include the error on the geometrical factor $\langle \beta \rangle$. Its value varies in a range $3\%-6\%$ depending on the cluster when the different CFHTLS Deep fields are used to estimate it. Moreover, in the present case, these 2 quantities are second order effects and the main source of
uncertainties remains the intrinsic ellipticity of the lensed galaxies (e.g. \citealt{hoekstra07}).

%__________________________________________________________________

\subsection{Comparison of the X-ray and weak lensing masses}
Raw  \xmm\ data were processed as described in \citet{pointecouteau05}
and \citet{pratt07}.  As for the \rexcess\ sample \citep{croston08},
surface brightness profiles were PSF-corrected, deprojected and converted
to 3D density profiles using the non parametric method described in
\citet{croston06}. 2D temperature profiles were extracted from \xmm\
spectroscopy as described in \citet{pratt10}. The 3D profiles were
recovered assuming analytical functions taking into account the PSF mixing
effects and weighting the contribution of temperatures in various rings. A
Monte Carlo procedure was used to compute the errors \citep{arnaud10}. The
hydrostatic mass equation was then applied to the density and temperature
profiles and to their logarithmic gradients to derive the observed mass
profile for each cluster.  Again a Monte Carlo method was applied to
determine the errors on each point, taking into account over-constraints
on the 3D profile imposed by the use of a parametric model. Observed
mass profiles were finally fitted with a NFW profile, constraining the
shape and the normalization parameters, i.e. $c_{500}$ and $M_{500}$.

\begin{figure}
\center
\rotatebox{0}{
\includegraphics[width=0.98\hsize]{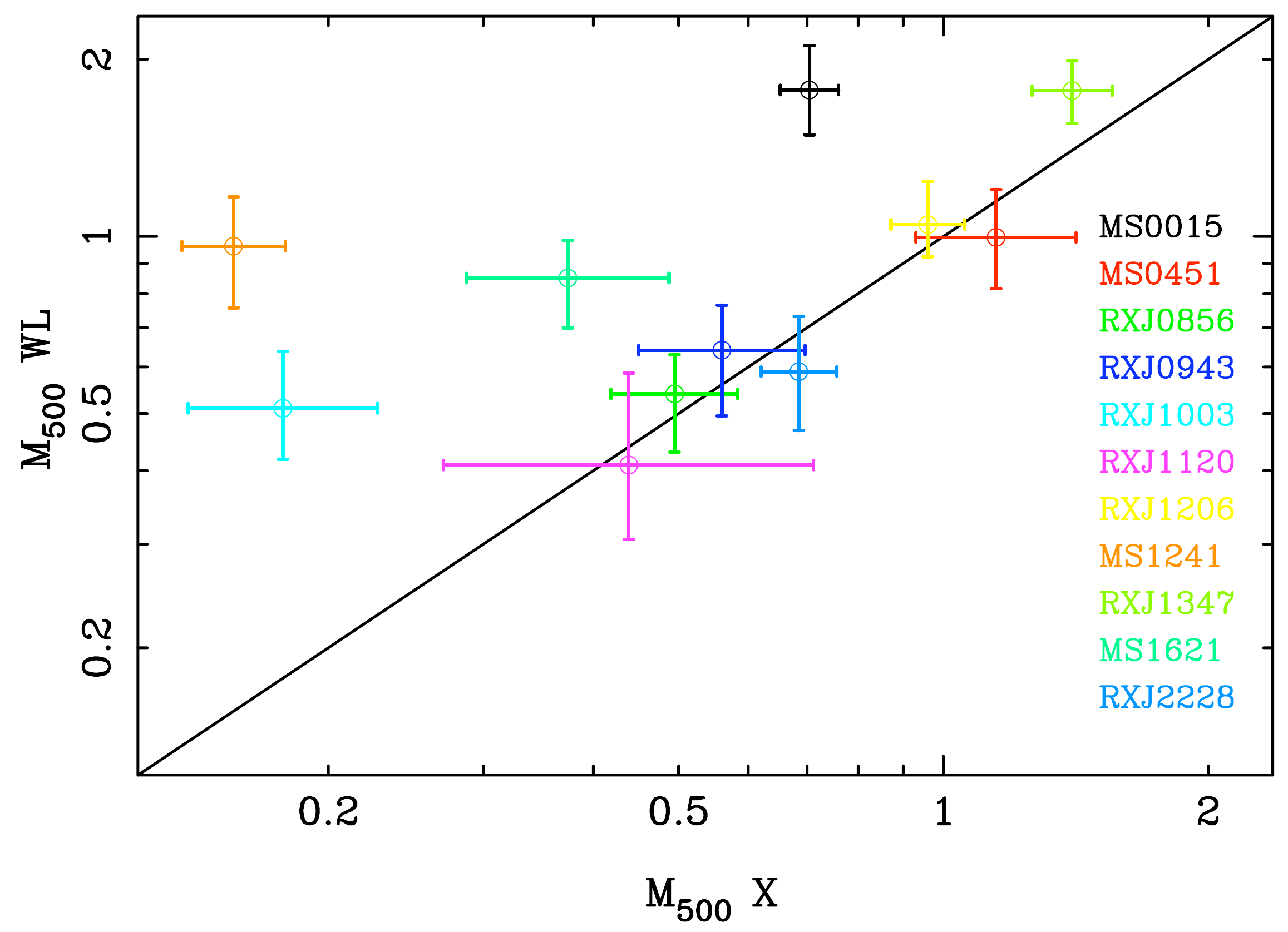}
}
\caption{Masses obtained with the weak lensing analysis (y-axis) versus
the hydrostatic equilibrium masses obtained with the X-ray analysis 
(x-axis). All masses are given in
units of $10^{15}h_{70}^{-1}M_{\odot}$ and are computed at the density
contrast $\Delta=500$. }
\label{fig:X_vs_WL}
\end{figure}

\begin{figure}
\center
\rotatebox{0}{
\includegraphics[width=0.98\hsize]{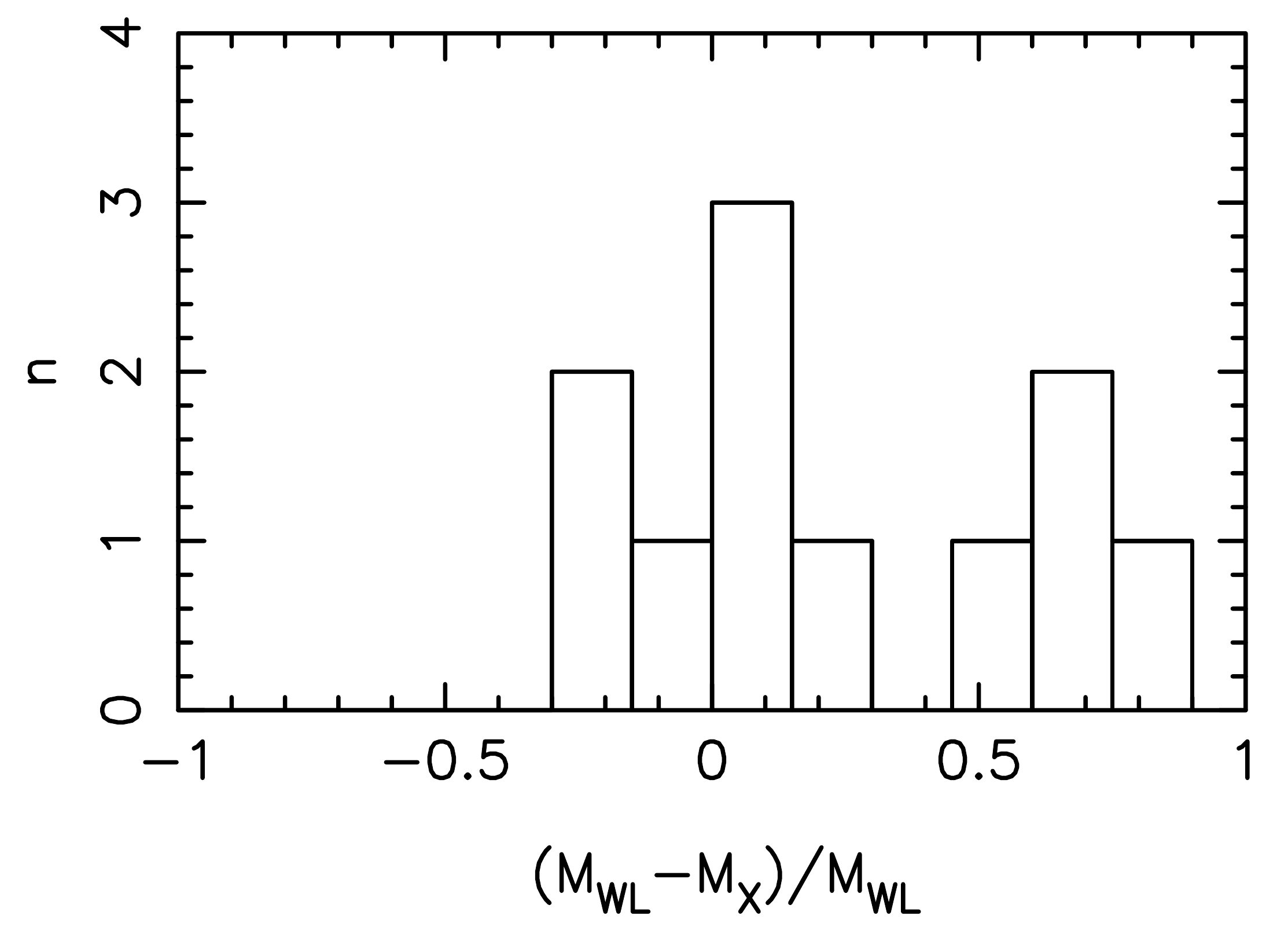}
}
\caption{Relative difference distribution between the weak lensing
masses and the X-ray masses for the 11 clusters of our sample. 
Masses are given at the density contrast $\Delta=500$.}
\label{fig:histo_X_WL}
\end{figure}

Figure \ref{fig:X_vs_WL} shows the comparison between the weak
lensing and the hydrostatic equilibrium masses derived from the
X-ray data for the 11 clusters. The comparison is made at a density contrast $\Delta=500$ to avoid any extrapolation of the X-ray data. The weak lensing masses $M_{200}$ are simply converted into $M_{500}$ assuming a concentration parameter $c_{200}=4$. This means that we are not comparing weak lensing and hydrostatic masses evaluated in the same physical radius. We are instead focusing here on a raw comparison between fitting results.\\
Over the whole sample, we almost reach a $1\sigma$ agreement between the 2 mass estimators since 7 clusters (i.e. {$\sim64\%$}) have compatible masses within their $1\sigma$ interval of individual uncertainties, with a small scatter and no strong
systematics, Fig. \ref{fig:histo_X_WL}. With the 11 clusters, we obtain an average mass ratio  $\langle M_{WL}^{500}/M_{X}^{500}\rangle = 1.92\pm1.58$. Removing MS 1241.5+1710 that is almost $3\sigma$ away from this average ratio, we find  $\langle M_{WL}/M_{X}\rangle = 1.49\pm0.75$, a value much less scattered. We recover here the 'classical' result that is a moderate excess for the weak lensing masses.

Similar studies were previously done on
clusters at lower redshifts \citep{zhang08,mahdavi08,zhang10}, but it is
the first time we show this kind of comparison at such a high redshift
range. Unfortunately, the weak lensing mass profiles are not sufficiently
constrained to allow a more detailed study of the variation with radius of the mass ratio $M_{WL}/M_{X}$. A combination with strong lensing information would be
needed to overcome this difficulty. It will be explored in a forthcoming
paper for the \excpres\ clusters for which strong lensing features are
detected (4 out of the 11 clusters).

For the 4 outliers identified in Figure \ref{fig:X_vs_WL}, their
individual properties will be presented in Paper 2. However it remains difficult to fully understand on a
case by case why these clusters show such discrepancies. It certainly
results from the mixed effect of intrinsic physical departures from
hydrostatic equilibrium and the limited quality of the lensing
data. Some differences are also statistically expected in unbiased
samples of clusters and can be characterized with numerical simulations
\citep{meneghetti10}. However, because of the limited size of the sample, we did not separate dynamically perturbed clusters from the relaxed one and try to highlight variations in the normalization and scatter of the $M_{WL}-M_{X}$ relation.\\

%__________________________________________________________________
\section{Mass-observable correlations}

\subsection{Methodology}
Because both masses and observables have uncertainties with possible
correlations (e.g. the richness depends on $R_{200}$ and thus on
the total mass), a simple $\chi^{2}$ minimization cannot be used to
calibrate scaling relations. Moreover, we expect to have an intrinsic
dispersion around the main scaling relation which needs to be included
and evaluated. So we preferred to perform a linear regression using the orthogonal BCES estimators ({\it Bivariate Correlated Errors and intrinsic
Scatter}, \citealt{akritas96}). This approach has already been used by
several groups for the calibration of mass-observable scaling relations
\citep{morandi07,pratt09}, expected to be characterized by power laws and so fittied by linear relations $y=ax+b$ in the log-log plan. For each
relation, both variables are normalized by a pivot close to their mean. This insures that the logarithmic slope and normalization of
the relation are nearly independent parameters.

Once the best fit of the correlation is obtained, the total dispersion over the $N$ points
\begin{equation}
\sigma_{tot}^{2}=\frac{1}{N-2}\sum_{i=1}^{N}(y_{i}-\alpha x_{i}-B)^{2}
\end{equation}
equals the quadratic sum of the statistical dispersion $\sigma_{stat}$ and the intrinsic one $\sigma_{int}$, which thus can be expressed as:
\begin{equation}
\sigma_{int}^{2}=\frac{1}{N-2}\sum_{i=1}^{N}\left[(y_{i}-\alpha x_{i}-B)^{2}-\frac{N-2}{N}(\sigma_{y_{i}}^{2}+\alpha^{2}\sigma_{x_{i}}^{2})\right]
\end{equation}

For each relation, we also estimated the Spearman correlation 
coefficient $\rho$:
\begin{equation}
\rho=1-\frac{6\sum d_{i}^{2}}{n(n^{2}-1)}
\end{equation}
where $d_{i}$ is the ranking difference of the {\it i-th} sorted mass
and observable vector. $\rho$ is a ranking correlation that shows the
monotonic degree of the relation between the two variables ($\rho=-1$ for
a strictly decreasing relation, $\rho=+1$ in the opposite case). Compared
to the classical Pearson coefficient, the Spearman coefficient does not
depend on the slope of the correlation.  Such a coefficient has been
used in similar studies to quantify the strength of a correlation,
e.g. \cite{lin04a}.

\subsection{Optical scaling laws}
%We shortly remind here how cluster galaxies are selected and how the
%total luminosity is estimated. More details will be given in Paper 2 which
%summarizes the optical properties of the clusters. First we select
%cluster galaxies as the population of bright galaxies belonging to the 
%red sequence of the color-magnitude diagram. The total luminosity
%$L_{200}$ and the optical richness $N_{200}$ are computed within
%the virial radius measured by the present weak lensing analysis. 

\begin{figure}
\center
\rotatebox{0}{
\includegraphics[width=0.98\hsize]{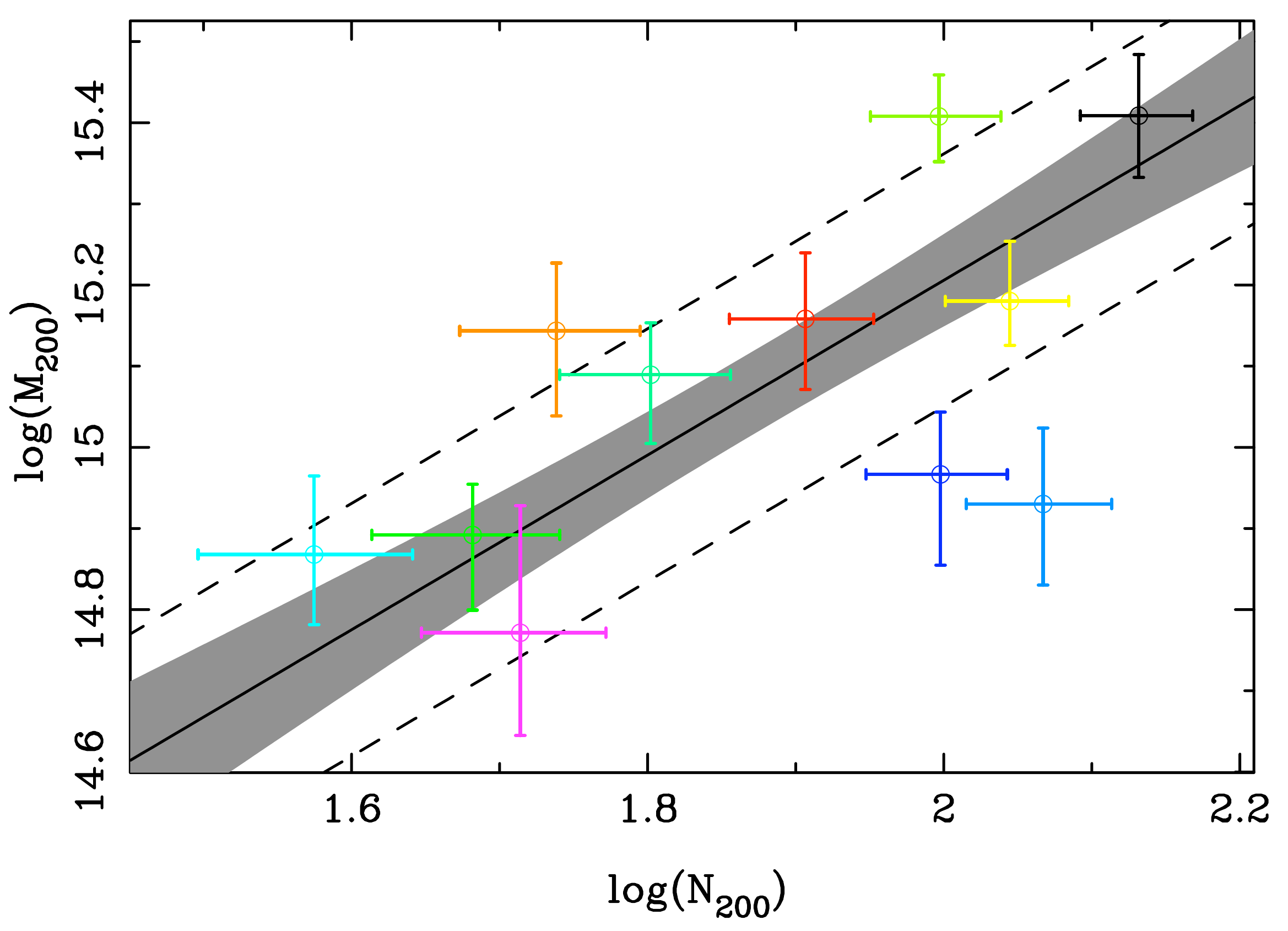}
}
\caption{Mass-richness scaling relation calibrated with the NFW lensing
masses $M_{200}$. The
black line shows the best BCES orthogonal fit. The region delimited by
the two dashed lines corresponds to the intrinsic dispersion of the
galaxy clusters around the best fit. The shaded area gives the
statistical $1\sigma$ uncertainty given by the best fit parameters.}
\label{fig:MnWL}
\end{figure}

First, we compared the weak lensing
mass $M_{200}$ to the optical richness $N_{200}$. We define $N_{200}$ as number of galaxies within $R_{200}$ that belong to the cluster red sequence (the precise definition will be presented in Paper 2). We keep only the brightest members with $L<0.4L^\star$. $L^\star$ is derived from the values of \cite{ilbert05} and according to the cluster redshifts. Richnesses are corrected from the background contamination but are not corrected from any luminosity cut nor any color selection. The aim is to remain consistent with previous studies that focus on the bright elliptical cluster galaxies.\\
The BCES calibration of this $M_{200}-N_{200}$
relation, given in Table \ref{table:scaling}, presents a
correlation with a logarithmic slope $\alpha=1.04\pm0.38$ (Fig.
\ref{fig:MnWL}). Despite a rather large uncertainty, the slope is
very close to 1, in agreement with the simplest model of structures
formation \citep{kravtsov04}. The uncertainties of the BCES fit are
dominated by the intrinsic dispersion (dashed lines) due to the limited
size of the cluster sample and increased by one outlier,  RX
J2228.5+2036. These results
are coherent with previous studies conducted at lower redshifts
\citep{becker07,johnston07,reyes08,rykoff08b,mandelbaum08,rozo09}.
We also compared the X-ray
masses with the cluster richnesses. This leads to a correlation with
a slope $\alpha=1.14\pm0.18$. In addition, we also looked at a possible correlation between the mass and
the richness evaluated in a given physical radius. In practice, we used
the richness determined within 1 Mpc (roughly equal to $0.5R_{200}$).
The BCES estimator gives a slope $\alpha=1.60\pm0.92$, a result with a
very large uncertainty that is useless with the present data.

To compare these different approaches, we computed the relative error
$R=| (M_{WL}-M_{proxy})/M_{WL}| $ averaged over the sample of
clusters after removing the main outlier RX J2228.5+2036. The $M-N_{200}$
relation leads to $R=0.27$ while with the $M-N_{1\,\mathrm{Mpc}}$
relation we find $R=0.41$. Even if the
conversion between richness and mass within a fixed aperture is less
physical than the use of the virial radius $R_{200}$, we show here
that it is still acceptable. Clearly the uncertainties in the scaling
relations are dominated by the intrinsic dispersion more than by the
choice of the working radius (see also \citealt{andreon10}). The accurate determination of $R_{200}$
is a second order effect and the use of a simple fixed aperture for
the measure of the cluster richness can already lead to a correct mass
proxy. This is particularly true as our sample covers a restricted
interval in both mass and redshift, so the fixed aperture is not far
from a fixed density contrast.

\begin{table*}
\caption{Summary of the fitting results for the mass-observable scaling
relations. Columns are (1) scaling relation; (2) best fit logarithmic slope; (3) best fit normalization; (4-5-6) total-statistic-intrinsic logarithmic dipsersions; (7) Spearman coefficient. Masses are expressed in units of $10^{15} \ h_{70}^{-1}
\ M_{\odot}$ (except for the $L_{X}-M$ relation with masses in units of
$5\times 10^{14}\ h_{70}^{-1} \ M_{\odot}$). The optical luminosities are
given in units of $7\times10^{12} \ h_{70}^{-2} \ L_{\odot}$, the richness
is normalized to 80 and the temperature to 5 keV. X-ray luminosities
are given in units of $10^{45}\ h_{70}^{-2} \ \mathrm{erg \ s^{-1}}$.  
}

\label{table:scaling}
\centering
\begin{tabular}{l c c c c c c}
\hline\hline\noalign{\smallskip}
scaling law & $\alpha$ & $A$ & $\sigma_{raw}$ & $\sigma_{stat}$ & $\sigma_{int}$ & $\rho$\\
\noalign{\smallskip}\hline\noalign{\smallskip}
$M_{200}^{WL}-N_{200}$ & $1.04\pm0.38$ & $1.27\pm0.12$ & $0.17\pm0.05$
& $0.11\pm0.01$ & $0.13\pm0.06$ & $0.66$\\
\noalign{\smallskip}
%\ccor{$M_{500}^{X}-N_{500}$ & $1.14\pm0.18$ & $2.03\pm1.21$ & $0.16$ &
%$0.13$ & $0.10$ & $0.70$}\\
%\noalign{\smallskip}\hline\noalign{\smallskip}
$M_{200}^{WL}-L_{200}$ & $0.95\pm0.37$ & $1.28\pm0.20$ & $0.18\pm0.04$
& $0.10\pm0.01$ & $0.14\pm0.05$ & $0.64$\\
%$M_{500}^{X}-L_{500}$ & $1.20\pm0.20$ & $0.32\pm0.04$ & $0.19$ & $0.13$ & $0.14$ & $0.68$\\
%\noalign{\smallskip}\hline\noalign{\smallskip}
\noalign{\smallskip}\hline\noalign{\smallskip}
$h(z)^{-1}L_{X}-h(z)M_{200}^{WL}$ & $2.06\pm0.44$ & $0.29\pm0.07$ &
$0.27\pm0.06$ & $0.16\pm0.03$ & $0.19\pm0.08$ & $0.85$\\
$h(z)M_{200}^{WL}-kT$ & $1.33\pm0.42$ & $1.03\pm0.20$ & $0.15\pm0.04$
& $0.09\pm0.01$ & $0.12\pm0.05$ & $0.73$\\
\noalign{\smallskip}\hline
\end{tabular}
\end{table*}

Second, we tested the relationship between mass and optical luminosity. The $M-L$ correlation
gives a roughly constant mass-to-light ratio as the BCES estimator returns a slope
$\alpha=0.95\pm0.37$. Note that to be consistent with other similar
works, we do not correct the total luminosities from incompleteness. Our
catalogs are limited to $0.4 L^\star$ and a correcting factor of $\sim1.6$
should be included to get the total luminosity, assuming a standard
Schechter luminosity function. This is done in Paper 2 which discusses
in details the M/L ratio of the clusters.\\
The correlation between X-ray masses and
optical luminosities gives $\alpha=1.20\pm0.20$, a value also compatible
with a constant mass-to-light ratio. Note that, as for the $M-N$ relation, the calibration with the X-ray masses gives a smaller intrinsic dispersion.
%\ccom{!!! a refaire avec les M500 et L500 a venir !!!}

Previous studies tend to find a slope in the $M-L$ relation slightly
larger than 1 \citep{lin04a,popesso05,bardeau07,reyes08}.  This tendency
is predicted by some of the semi-analytical models of galaxy formation, at
least for massive clusters \citep{marinoni02} in which the star formation
efficiency can be inhibited by the long cooling times of hot gas on
large-mass scales. Merging processes and galaxies interactions can also affect the luminosity of
galaxies and are more efficient in massive clusters. But because of the
small mass range covered by our cluster sample, it is difficult to
draw firm conclusions on any possible variation of the M/L ratio with
mass. It would be necessary to add galaxy groups in the sample, although
accurate mass determinations are quite difficult for low mass
structures.

Moreover, the complex method to determine the value of $R_{200}$
represents an issue to estimate $L_{200}$. So we repeated the same
procedure as described above and compared the masses with luminosities
integrated inside a 1 Mpc radius. We also calculated the average relative
errors on the mass. We obtain $R=0.47$ for masses derived from the
$M-L_{1\,\mathrm{Mpc}}$ relation while we get $R=0.33$ when we use the
$M-L_{200}$ relation. Again, the intrinsic dispersion is the main
source of uncertainties, the precise determination of $R_{200}$ being a second order effect. 

\subsection{X-ray scaling laws}

%The calibration of the scaling relations based on
%the X-ray properties of the galaxy clusters over the whole \excpres\
%sample will be presented in a forthcoming paper. 
We focus here only on scaling laws fitted with the weak lensing
masses taken at $\Delta=200$. The calibration of the $L_{X}-M$
relation has been the subject of many studies, mainly at low
redshifts.  Mass estimators are usually derived from X-ray data
\citep{markevitch98,arnaud02,reiprich02,popesso05,morandi07,
pratt09,vikhlinin09}. Several attempts to use
weak lensing masses were also proposed more recently
\citep{hoekstra07,bardeau07,rykoff08b,leauthaud10,okabe10b}.
Most these studies converge towards a power law that differs from
the relation expected in the hierarchical scenario of structures
formation where the X-ray luminosity scales with the mass as
$F_{z}^{-1}L_{X}\propto(F_{z}M_{tot})^{4/3}$. The $F_{z}$ factor
stands for the evolution of the scaling relation and reduces to $h(z)$, the
reduced Hubble constant, for models of structures formation based only on gravitation and assuming a fixed density contrast to derive masses \citep{voit05}.\\
%when masses are evaluated at a given density contrast, e.g. $\Delta=500$
%or 200, because we assume both self similarity in the shape and mass profile of the clusters and %a constant gas fraction.  
Several processes can affect the properties of a cluster, both generating
an intrinsic dispersion around the expected $L_{X}-M$ relation and a
breaking of self-similarity. Most of previous studies point towards a
slope that is slightly larger than the expected one, i.e. $\alpha=1.6-1.8$ instead of
4/3. The intrinsic dispersion is found to be quite large, up to almost 50\% \citep{reiprich02}. However, it appears
that when cool cores are removed, this intrinsic dispersion is smaller
\citep{allen98,markevitch98,voit02,maughan07,pratt09}. Indeed, the X-ray
luminosity is strongly affected by the physical mechanisms ruling the
baryonic content of clusters  (pre-heating, radiative cooling, feedback
of supernovae and active galactic nuclei ...), and so is the $L_X-M$ with
respect to the simplest gravitational model. These processes are modeled
in up-to-date numerical simulations that combine the gravitational
evolution of dark matter structures with the hydrodynamical behavior
of the intra-cluster gas \citep{borgani04,kay04, borgani08}. These
simulations also seem to favor a slope of the $L_X-M$ relation larger
than the canonical one.

\begin{figure*}
\center
\rotatebox{0}{
\includegraphics[width=0.98\hsize]{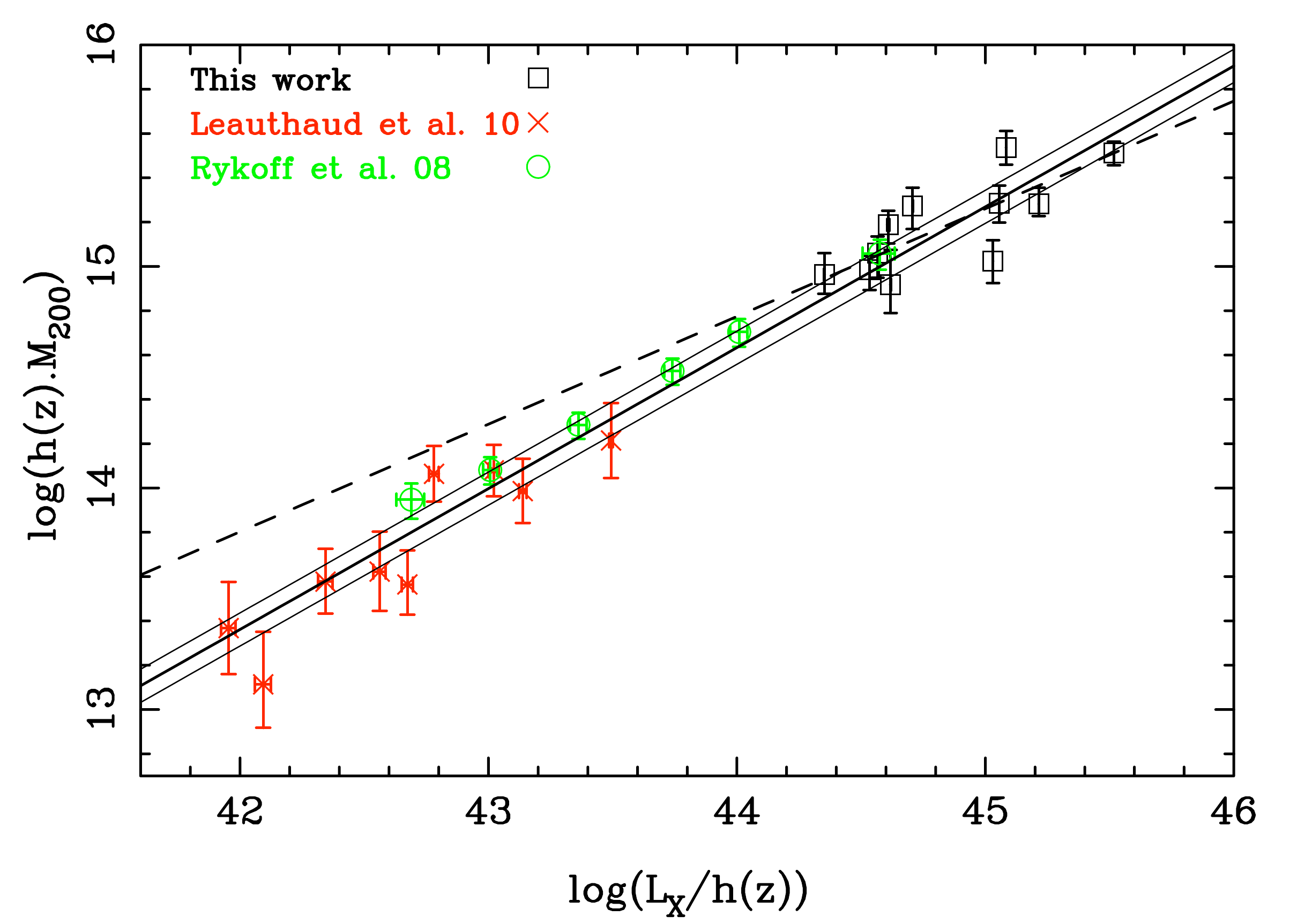}
}
\caption{Weak lensing masses versus X-ray luminosities using the
\excpres\ clusters (black boxes) combined with the stacked clusters and
groups of \cite{rykoff08b} (green circles) and of \cite{leauthaud10}
(red crosses). The solid lines show the best BCES fit and the 
corresponding intrinsic dispersion for the
three data sets. For comparison the BCES best fit obtained
with the \excpres\ weak lensing sub-sample only is plotted (dashed line).
}
\label{fig:LxMcompil}
\end{figure*}

The results of the BCES estimator for the $L_{X}-M_{200}^{WL}$
relation are presented in Table \ref{table:scaling}. Although our $L_X$ measurements do not purely
equate $L_{X, 200}$ (see Table \ref{table:clusters}), our best fit
relation is in good agreement with previous works as we find a slope
$\alpha=2.06\pm0.44$. Despite a strong correlation, we also observe a
large scatter around the best fit with $\sigma_{int}=0.19$.

These encouraging results should nonetheless be taken with caution as our
sub-sample of the \excpres\ clusters was designed to keep only the most
luminous clusters in the sample. It therefore covers a limited range in
X-ray luminosity. In order to explore a larger range, we combined this
sample with clusters from two studies: the COSMOS galaxy groups analyzed
in \cite{leauthaud10} and the maxBCG galaxy groups and clusters presented
in \cite{rykoff08b} (Fig.  \ref{fig:LxMcompil}). Both are using stacks
of objects with similar richness to improve the statistics of the shear
profiles for low mass objects. Their X-ray luminosities are also stacked
within $R_{200}$ apertures, whereas our values of the luminosities are
computed within the detection radius (see Table \ref{table:clusters}).
However, we checked that the ratio between $R_{det}$ and $R_{200}$ for
our sample is $\sim 0.9$ on average. We therefore consider that the
present values are close enough to the values within $R_{200}$, thus
allowing us to compare and combine them with the aforementioned works.
%Values of $L_X$ within a fixed density constrast will be reported in Arnaud et al. (in preparation).  
This combination of the three data sets allows to cover an extended
range in mass and luminosity. We fit them simultaneously and the best
fit relation gives a slope of $\alpha=1.57\pm0.07$. This slope is lower
than that from our data only but still in disagreement with the hierarchical prediction of 4/3. The intrinsic dispersion stays roughly
unchanged over the combined sample.

%\begin{figure}
%\center
%\rotatebox{0}{
%\includegraphics[width=0.98\hsize]{M200TWL}
%}
%\caption{Same as Figure \ref{fig:MLxWL} for the mass-temperature
%scaling relation. Temperatures are expressed in keV. The red line
%corresponds to the best fit with a slope of $1.5$. Masses are corrected
%from evolution with the $h(z)$ factor.}
%\label{fig:MTWL}
%\end{figure}
Although it is not nowadays much used as a mass proxy, we then checked
the $M_{WL}-T$ relation.
% using the global X-ray temperature provided
%in Table~\ref{table:clusters}. Even if not excised, these temperature
%are very consistent with those extracted within $[0.15-0.75]~R_{500}$.
%The average ratio over the sample is $0.98\pm0.05$. 
We derived a slope of $\alpha=1.33\pm0.42$, a value compatible with the
self-similar prediction of 1.5 within the $1\sigma$ limit. This weak lensing calibration is compatible with those
by \cite{hoekstra07} and \cite{okabe10b}, who respectively derive a
slope of $\alpha=1.34\pm0.29$ and  $\alpha=1.49\pm0.58$.

We also computed the average relative error $R$ between the weak lensing
masses and the mass proxies determined with X-ray data. The $L_{X}-M$
relation gives $R=0.28$ and the $M-T$ lead to $R=0.31$. These two values are of the same
order of those previously obtained for the $M-N$ and $M-L$ correlations
with optical data. 

\section{Summary and conclusions}
In this paper we provide for the first time a thorough weak lensing
analysis of a sample of 11 galaxy clusters (a sub-sample of the \excpres\
sample -- Arnaud et al. in preparation) in a relatively high redshift
range, $0.4<z<0.6$.  We have conducted a careful validation of our
weak lensing pipeline, thanks in particular to the STEP simulations
which allowed to re-calibrate our shear measurements. The shear
profiles were fitted with a NFW profile with a fixed concentration
parameter $c_{200}=4$. This is due to the lack of good constraints
on the shear profile near the center so only the total mass $M_{200}$
is a reliable measure.  These weak lensing masses were compared to the
X-ray masses derived from  \xmm\ data. Over the 11 clusters, we find
a good match between these two mass estimators for 7 clusters while we
found 4 outliers, thus having almost a $1\sigma$ agreement in the two methods. Such results are consistent with other similar studies
performed for lower redshift samples.

The primary goal of this study was to look for correlations between
the total mass of galaxy clusters and several baryonic tracers.
In our sample of massive clusters, we found that the total mass is
proportional to the optical richness of galaxies.  This result
is in good agreement with previous works at lower redshifts
\citep{becker07,johnston07,reyes08,rykoff08b,mandelbaum08,rozo09}.
We reach the same conclusion for the correlation between mass and optical
luminosity, with a constant ratio $M/L$, even if the correlation is weaker
with a larger intrinsic dispersion.  Using \xmm\ data and the weak lensing
masses, we build scaling relations with the X-ray total luminosity and
the global spectroscopic temperature. Both correlations, although weakly
constrained due to limited coverage of the mass range, give results in
good agreement with previous studies: (i) the mass-luminosity relation
presents a non self-similar slope, larger than expected from a purely
gravitational model of structure formation. (ii) the mass-temperature
relation is roughly compatible with the hierarchical prediction. We
have extended our investigation of the $L_X-M$ relation by combining
our cluster sample with two samples of groups and clusters for which
weak lensing masses were obtained \citep{rykoff08b,leauthaud10}. The
resulting $L_X-M_{200}$ relation covers nearly two decades in mass and
displays a regular shape with a well constrained slope.

Even if the limited size of our sample and its reduced mass range coverage
allow us to only probe the high end of the mass distribution of clusters,
we have demonstrated the synergy with the X-ray measurements (i.e., mass,
luminosity and temperature). More specifically, we have confirmed that
the scaling relations with the lensing mass proxy are already in place at
$z \sim 0.5$, with, at first order, no significant departures from the
same relations at lower redshift (assuming a self-similar evolution).
The limited quality of the lensing data for our sample restricts
the added value of a joint analysis, but the weak lensing and X-ray
coverage of cluster prove to be complementary. Indeed X-ray data allow,
to efficiently probe the central regions out to about $R_{500}$. Beyond
this radius the X-ray brightness rapidly drops, whereas the weak-lensing
signal picks up to characterize the outer parts of massive halos.
A perspective work would be to investigate the strong lensing signal
(see Table\ref{table:clusters}) of each cluster in our sample, in order
to perform a full lensing analysis from the inner part (with the strong
lensing signal) to the outer parts (with the weak lensing signal). This
would provide a coherent mass proxy estimator to be compared directly
with X-ray and dynamical proxies.

Finally, in the perspective of large optical surveys such as the Euclid
space mission, self calibrated mass proxies will be needed for the
full scientific exploitation of the mission. The $M-N$ is an obvious
candidate, which tight calibration will require validation against
other relations and mass proxies, such as the X-ray mass. The $M-N$
relation we provide in this paper is an early result in the redshift range
$0.4<z<0.6$. It illustrates the difficulty to constrain this particular
scaling relation. Provided the $N$ estimator is made as universal as
possible, which can be a non trivial task, this relation is promising
in the framework of optical surveys.

\begin{acknowledgements}
We wish to thank Roser Pello for many fruitful discussions. She provided
the catalogs of photometric redshifts built from the CFHTLS Deep fields
which are used in this work. We are grateful to the TERAPIX team who
processed part of the data efficiently. We also thank the Programme
National de Cosmologie et Galaxies of the CNRS for financial support.
ML acknowledges the Centre National de la Recherche Scientifique (CNRS) for its support.
The Dark Cosmology Centre is funded by the Danish National Research Foundation.
The present work is based on observations obtained with  \xmm\,
an ESA science mission with  instruments and contributions directly
funded by ESA Member States and the USA (NASA).  This research used the
facilities of the Canadian Astronomy Data Centre operated by the National
Research Council of Canada with the support of the Canadian Space Agency.
This work is based on observations obtained with MegaPrime/MegaCam,
a joint project of CFHT and CEA/DAPNIA, at the Canada-France-Hawaii
Telescope (CFHT) which is operated by the National Research Council
(NRC) of Canada, the Institut National des Science de l'Univers of the
Centre National de la Recherche Scientifique (CNRS) of France, and the
University of Hawaii.  
\end{acknowledgements}

%
%_____________________________________________________________

\bibliography{references}

\clearpage

\begin{center}
\section*{Annexes}
\end{center}

\begin{figure}[!ht]
\centering
\rotatebox{0}{
\includegraphics[width=0.98\hsize]{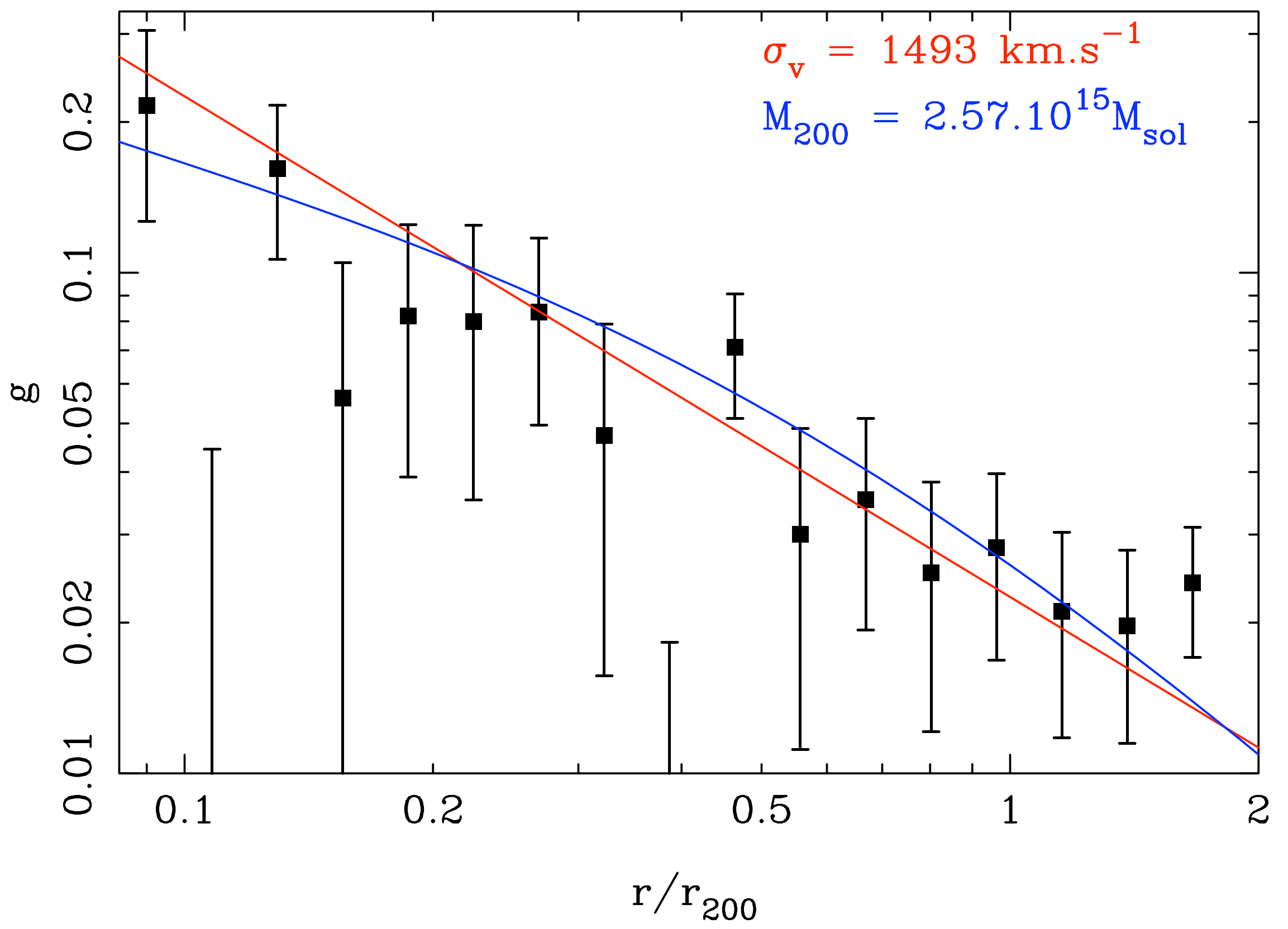}
}
\caption{Reduced shear profile normalized by the virial radius
$g(r/R_{200})$ for the galaxy cluster MS 0015.9+1609. The velocity
dispersion $\sigma_{v}$ and the total mass $M_{200}$ fitted to 
this profile are given (red curve for the SIS model, blue one for the NFW
model).}
\label{fig:MS0015sprof}
\end{figure}

\begin{figure}[!ht]
\centering
\rotatebox{0}{
\includegraphics[width=0.98\hsize]{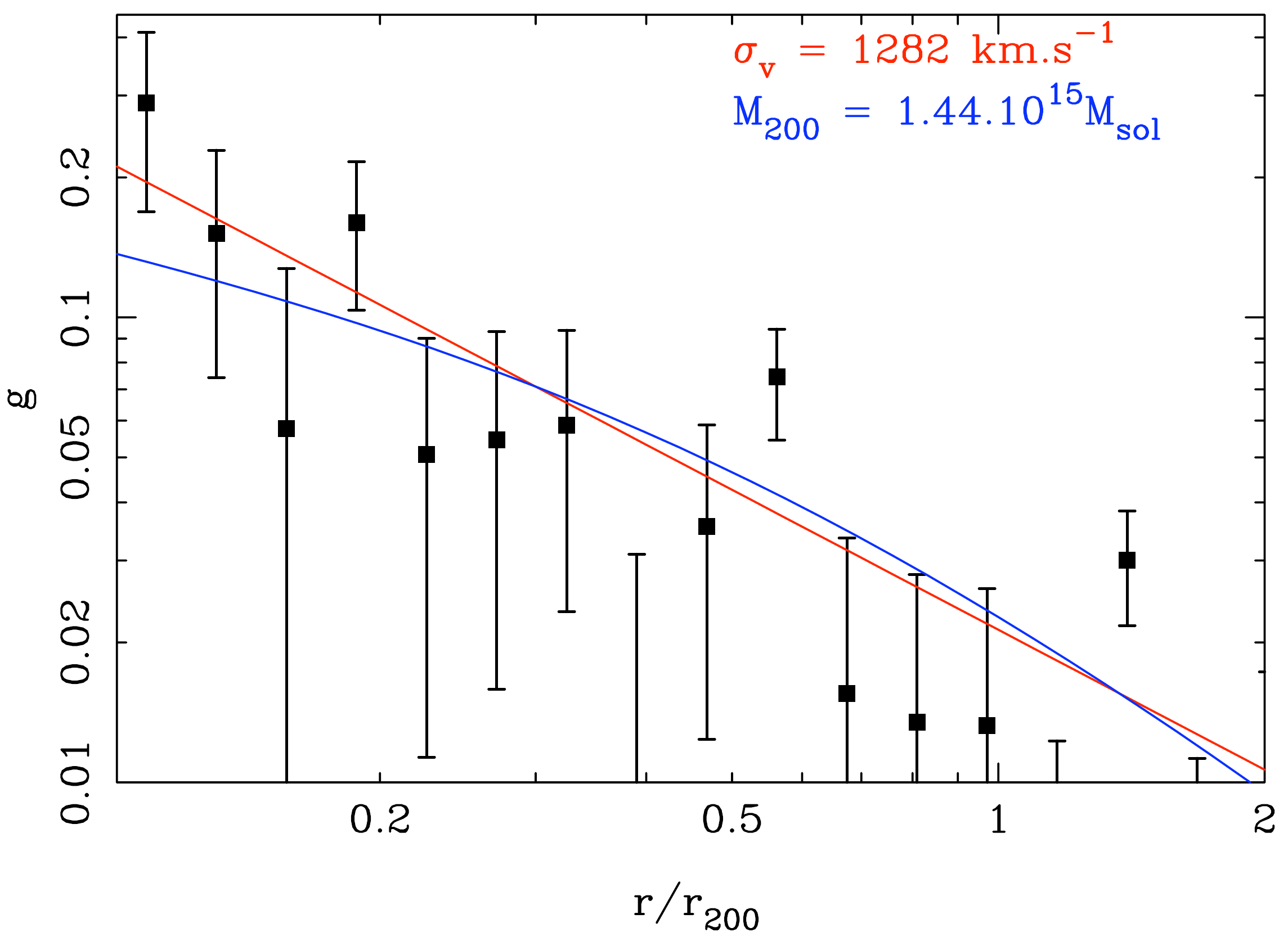}
}
\caption{Same as Figure \ref{fig:MS0015sprof} for the cluster MS 0451.6--0305.}
\label{fig:MS0451sprof}
\end{figure}

\begin{figure}[!ht]
\centering
\rotatebox{0}{
\includegraphics[width=0.98\hsize]{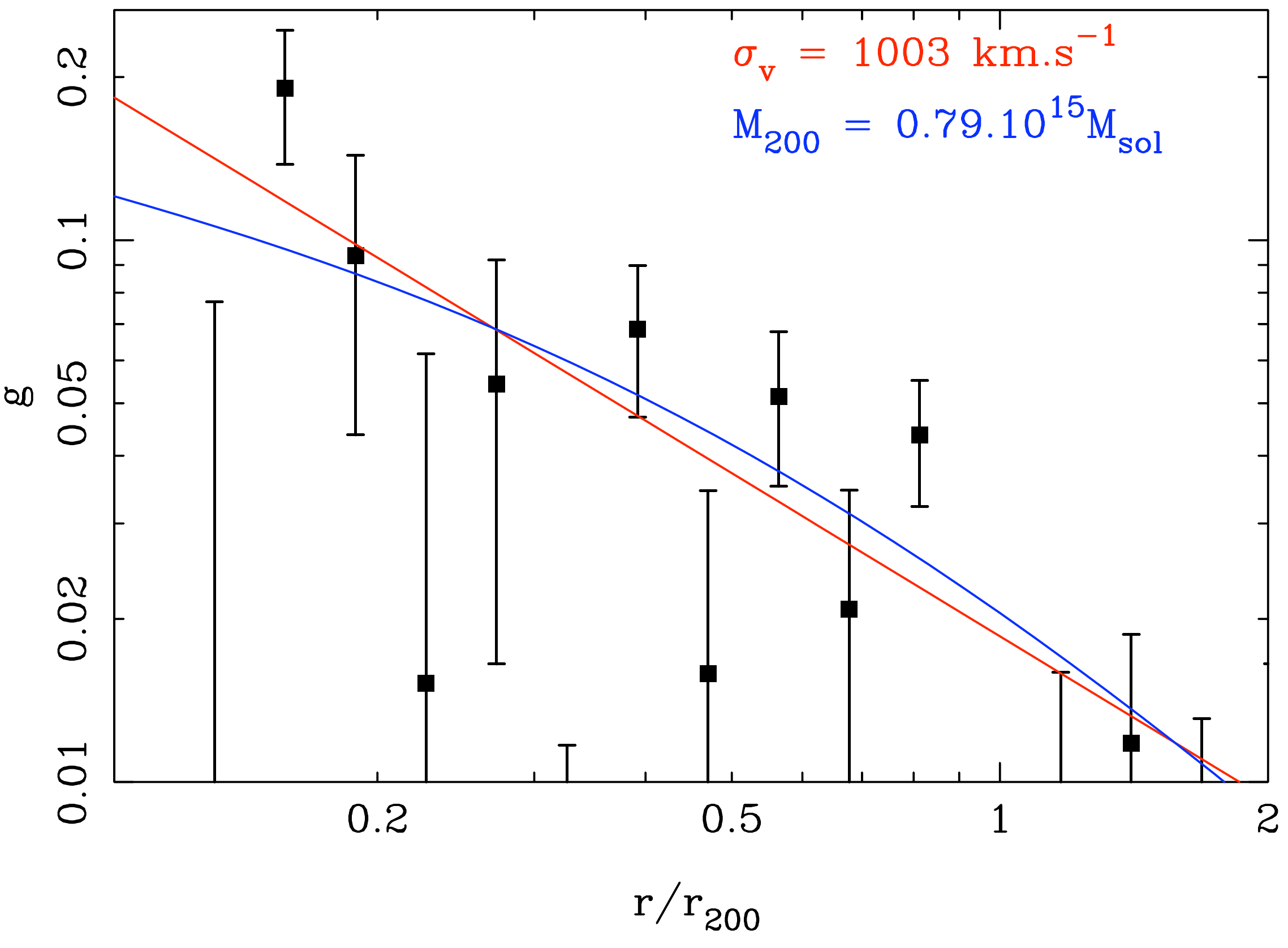}
}
\caption{Same as Figure \ref{fig:MS0015sprof} for the cluster RXC J0856.1+3756.}
\label{fig:RXJ0856sprof}
\end{figure}

\begin{figure}[!ht]
\centering
\rotatebox{0}{
\includegraphics[width=0.98\hsize]{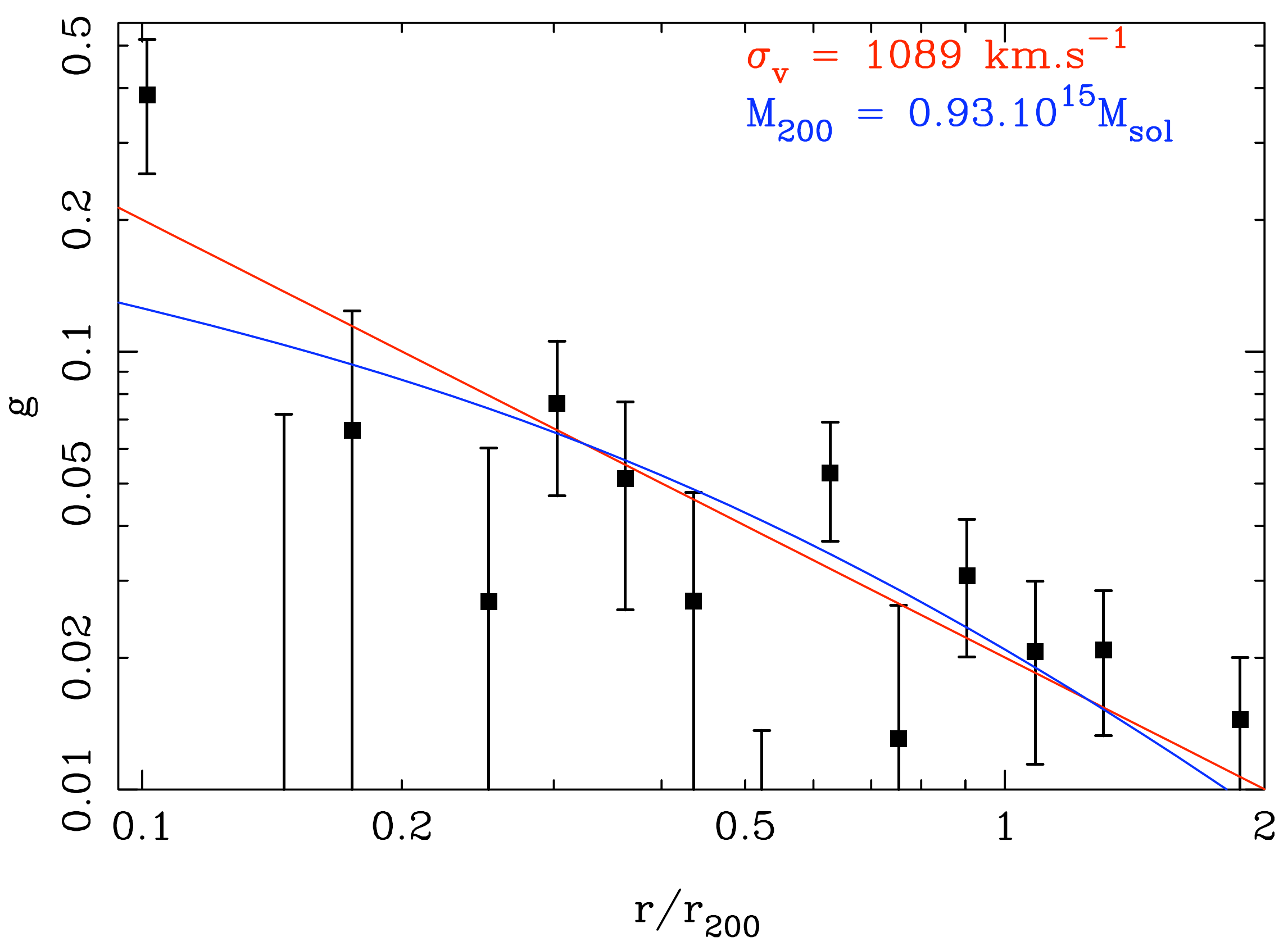}
}
\caption{Same as Figure \ref{fig:MS0015sprof} for the cluster RX J0943.0+4659.}
\label{fig:RXJ0943sprof}
\end{figure}

\begin{figure}[!ht]
\centering
\rotatebox{0}{
\includegraphics[width=0.98\hsize]{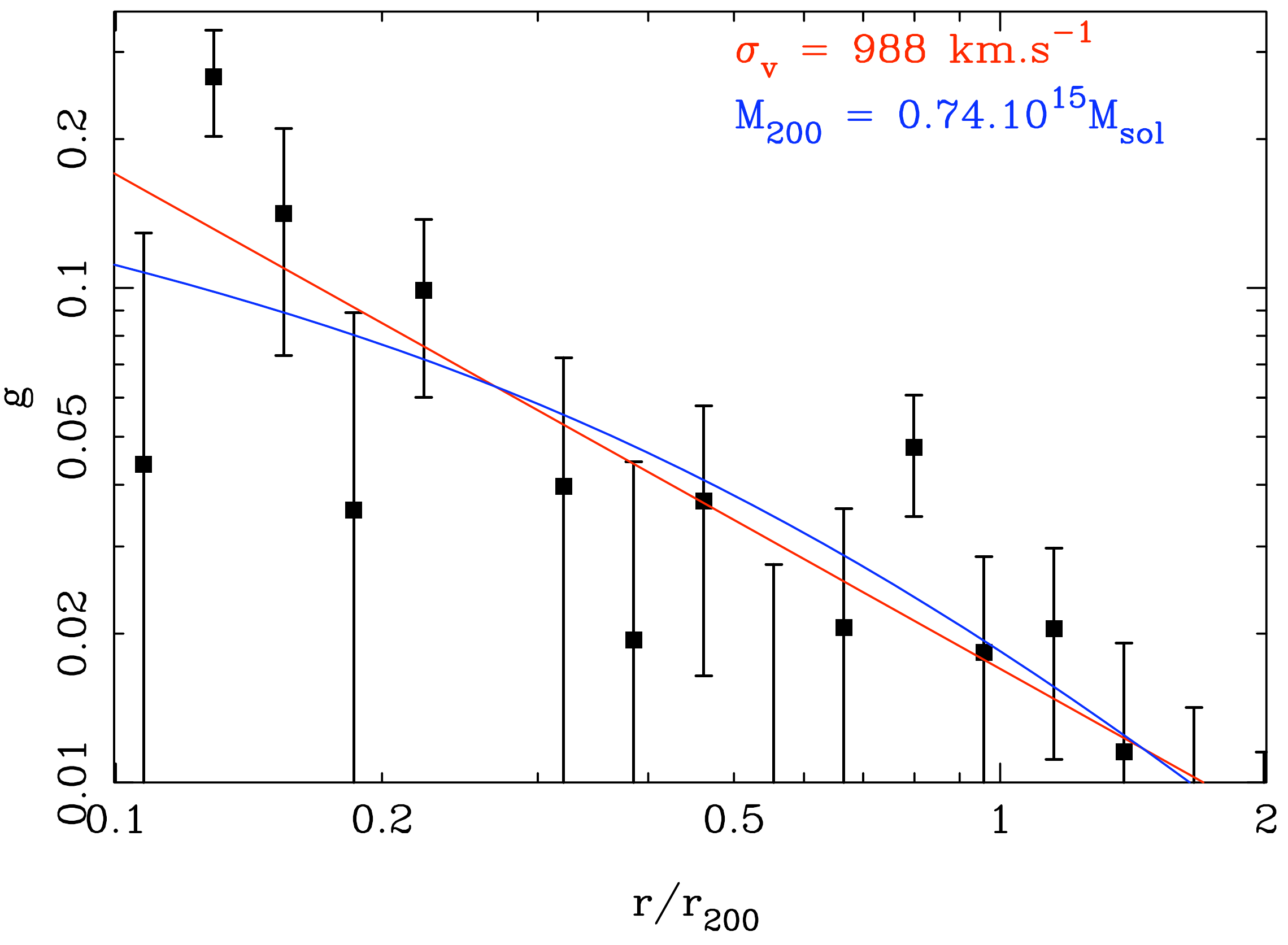}
}
\caption{Same as Figure \ref{fig:MS0015sprof} for the cluster RXC J1003.0+3254.}
\label{fig:RXJ1003sprof}
\end{figure}

\begin{figure}[!ht]
\centering
\rotatebox{0}{
\includegraphics[width=0.98\hsize]{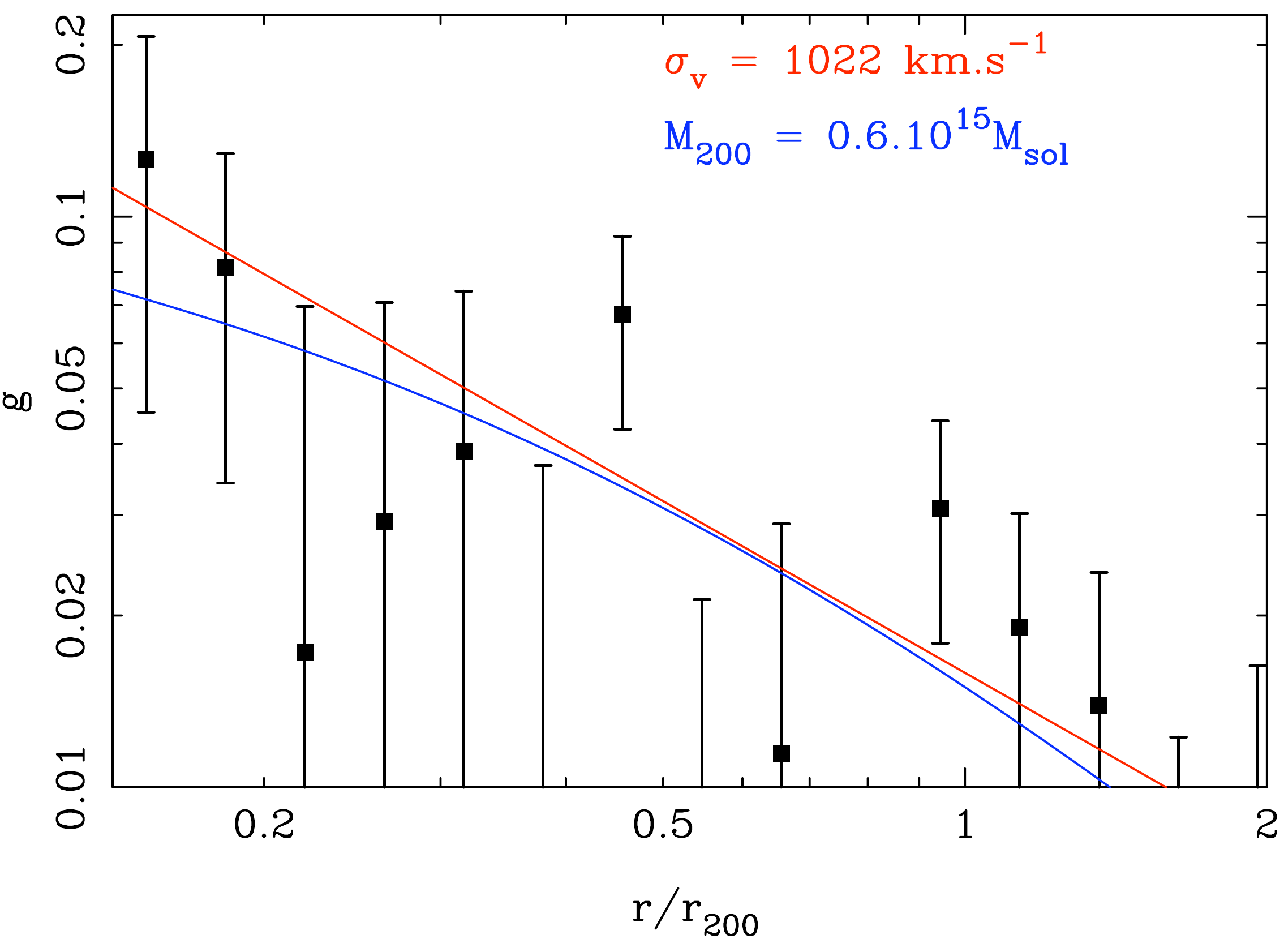}
}
\caption{Same as Figure \ref{fig:MS0015sprof} for the cluster RX J1120.1+4318.}
\label{fig:RXJ1120sprof}
\end{figure}

\begin{figure}[!ht]
\centering
\rotatebox{0}{
\includegraphics[width=0.98\hsize]{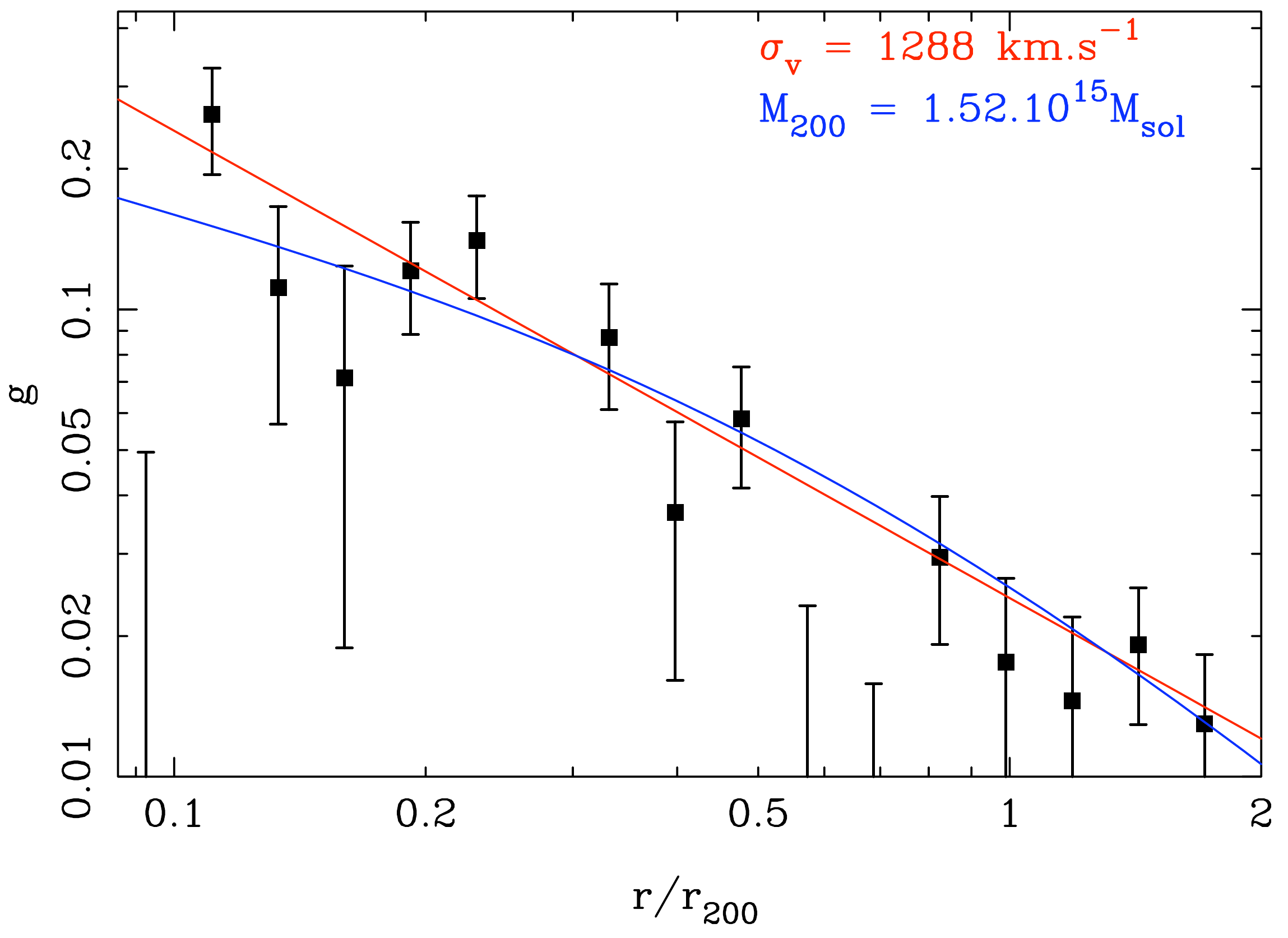}
}
\caption{Same as Figure \ref{fig:MS0015sprof} for the cluster RXC J1206.2--0848.}
\label{fig:RXJ1206sprof}
\end{figure}

\begin{figure}[!ht]
\centering
\rotatebox{0}{
\includegraphics[width=0.98\hsize]{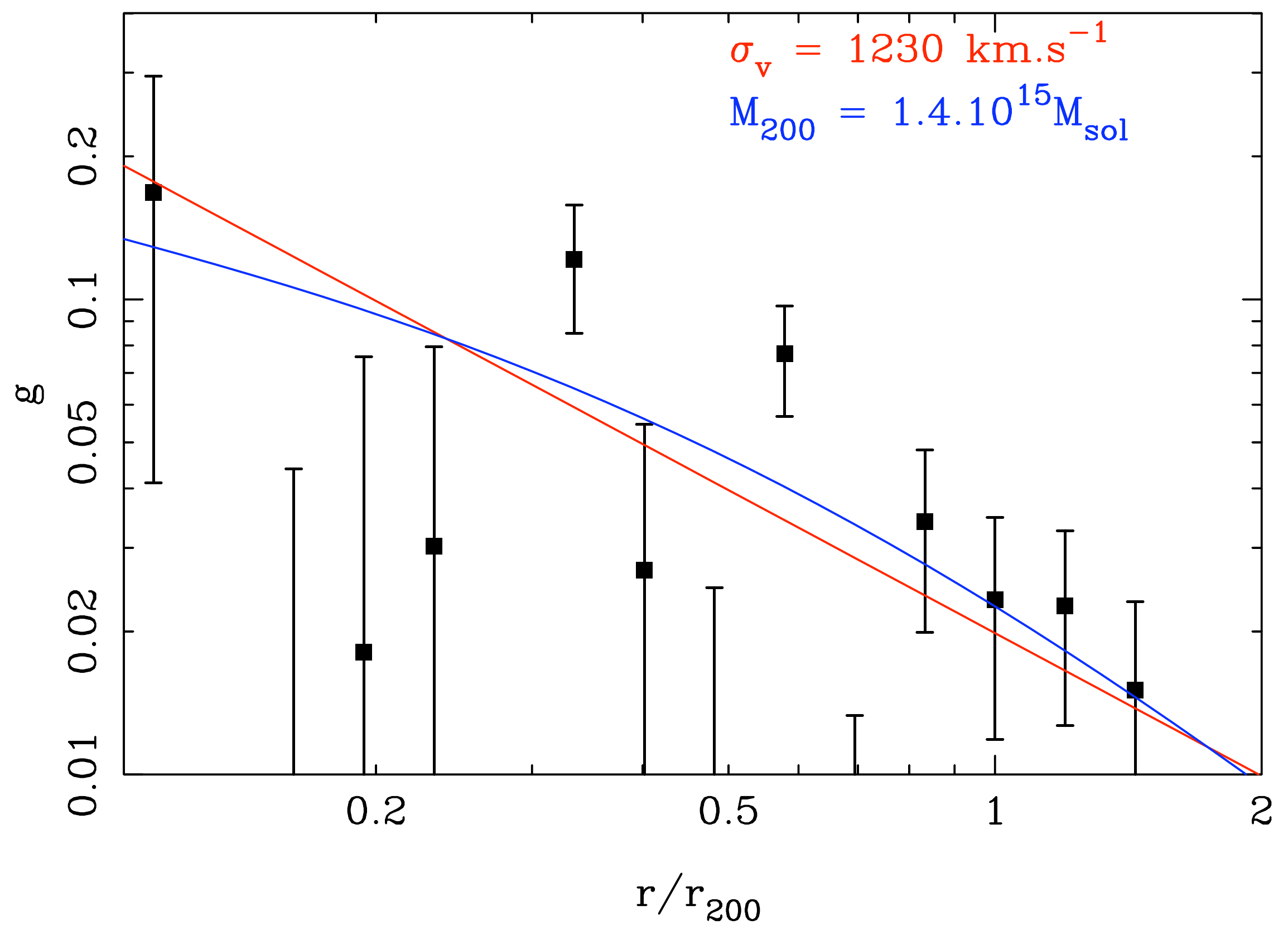}
}
\caption{Same as Figure \ref{fig:MS0015sprof} for the cluster MS 1241.5+1710.}
\label{fig:MS1241sprof}
\end{figure}

\begin{figure}[!ht]
\centering
\rotatebox{0}{
\includegraphics[width=0.98\hsize]{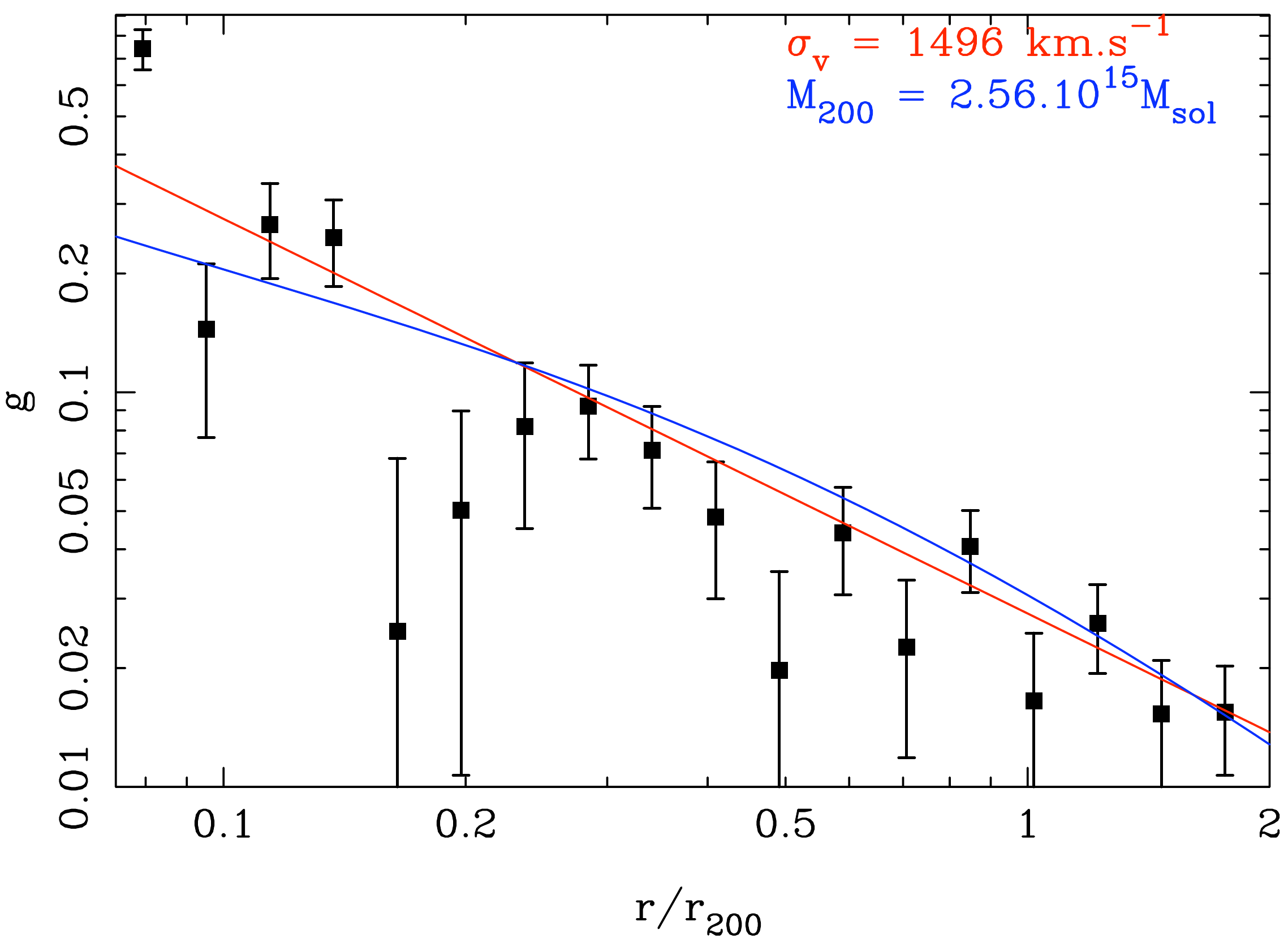}
}
\caption{Same as Figure \ref{fig:MS0015sprof} for the cluster RX J1347.5--1144.}
\label{fig:RXJ1347sprof}
\end{figure}

\begin{figure}[!ht]
\centering
\rotatebox{0}{
\includegraphics[width=0.98\hsize]{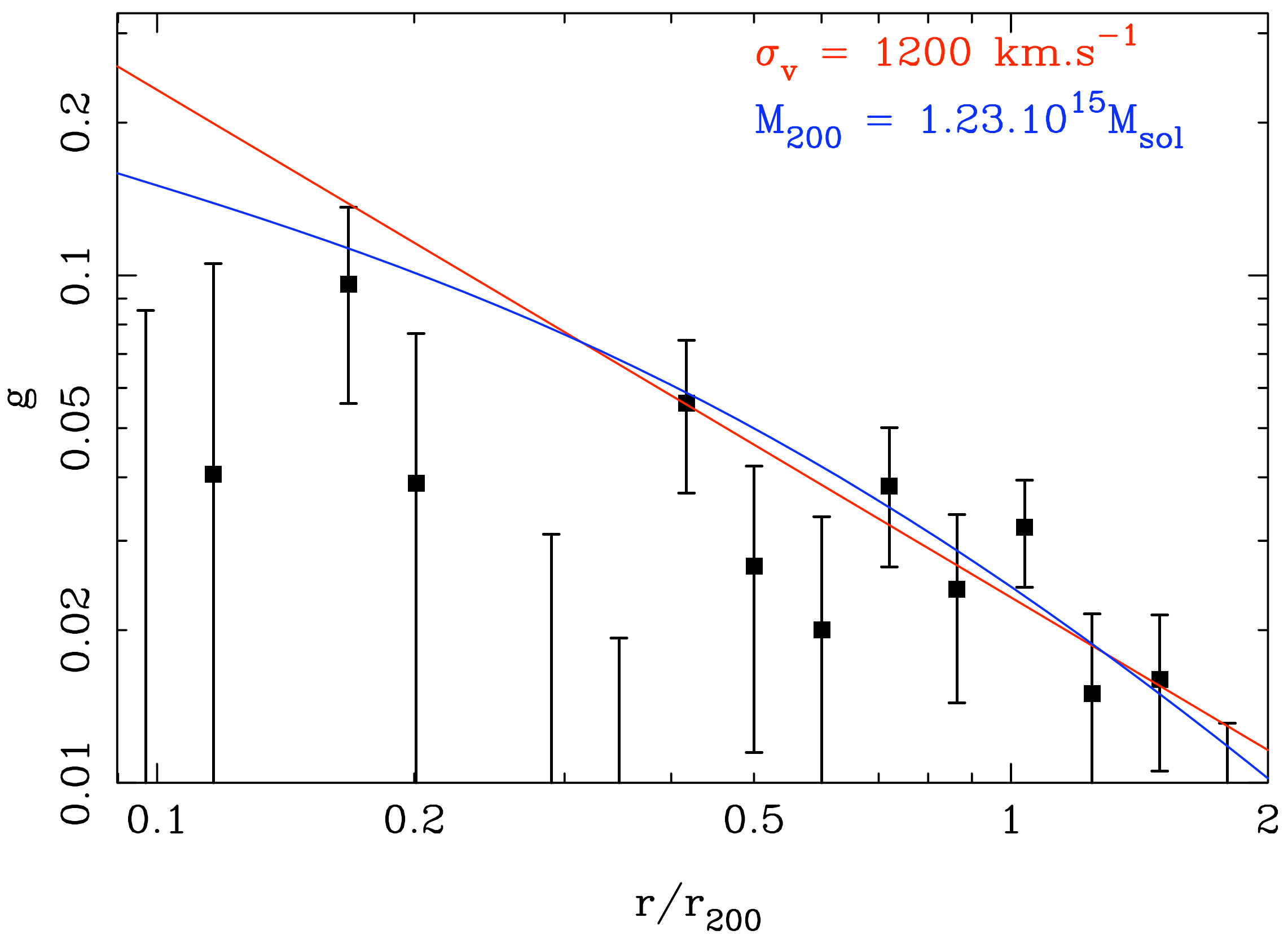}
}
\caption{Same as Figure \ref{fig:MS0015sprof} for the cluster MS 1621.5+2640.}
\label{fig:MS1621sprof}
\end{figure}

\begin{figure}[!ht]
\centering
\rotatebox{0}{
\includegraphics[width=0.98\hsize]{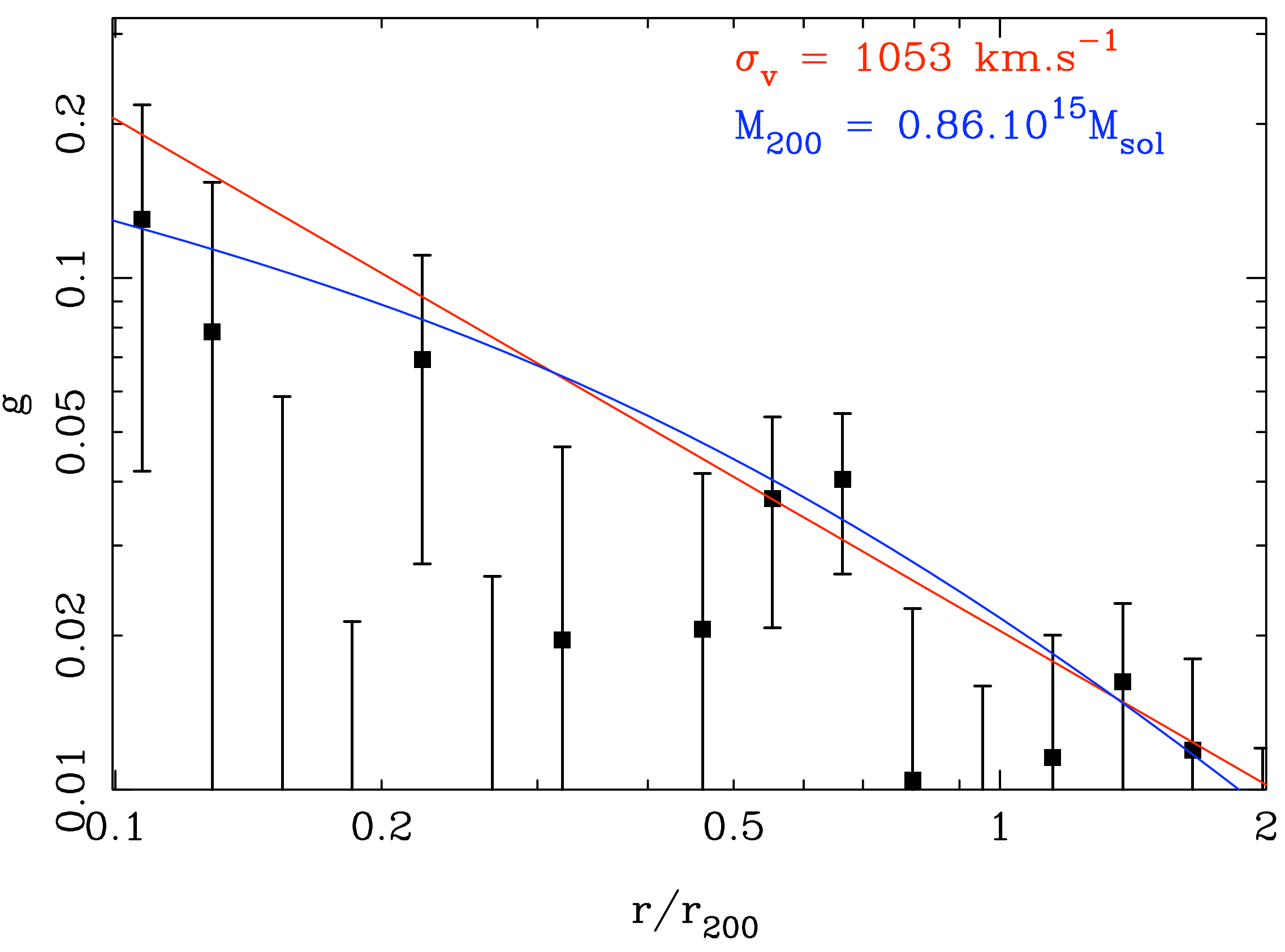}
}
\caption{Same as Figure \ref{fig:MS0015sprof} for the cluster RX J2228.5+2036.}
\label{fig:RXJ2228sprof}
\end{figure}

\end{document}